\documentclass[11pt,a4paper,twoside,groupcitations]{article}
\usepackage[T1]{fontenc}
\usepackage[ansinew]{inputenc}
\usepackage[english]{babel}
\usepackage{amsfonts}
\usepackage{amsmath}
\usepackage{bm}
\usepackage{array}
\usepackage{amsthm}
\usepackage{amssymb}
\usepackage{graphicx}
\usepackage{subfigure}
\usepackage{braket}
\usepackage{eucal}
\usepackage{verbatim}
\usepackage[table]{xcolor}
\usepackage{caption}
\usepackage{cite}
\usepackage{textcomp}
\usepackage{amssymb,mathrsfs,physics}
\usepackage{comment}
\usepackage{multicol}
\usepackage{tikz}
\usetikzlibrary{positioning,arrows}
\usetikzlibrary{decorations.pathmorphing}
\usetikzlibrary{decorations.markings}
\usetikzlibrary{calc,decorations.markings}
\usetikzlibrary{arrows,shapes}
\usetikzlibrary{matrix,arrows}
\usepackage{pgfplots}
\usepackage{xparse}
\definecolor{jade}{HTML}{00A86B}
\newcommand{\be}{\begin{eqnarray}}
\newcommand{\ee}{\end{eqnarray}}

\renewcommand{\d}{\mbox{${\rm d}$}} %d differenziale non corsivo in math mode

\newcommand{\gn}{G_{\rm N}}
\newcommand{\rh}{r_{\rm H}}
\newcommand{\Rh}{R_{\rm H}}

\title{\bf Effective metric outside bootstrapped Newtonian sources}
\author{Roberto~Casadio$^{ab}$\thanks{E-mail: casadio@bo.infn.it},
$\ $
Andrea~Giusti$^{c}$\thanks{E-mail: agiusti@ubishops.ca},
$\ $
Iber\^e~Kuntz$^{ab}$\thanks{E-mail: kuntz@bo.infn.it},
$\ $
and 
Giulio~Neri$^{a}$\thanks{E-mail: giulio.neri3@studio.unibo.it}
\\
\\
$^a${\em Dipartimento di Fisica e Astronomia, Universit\`a di Bologna}
\\
{\em via Irnerio~46, 40126 Bologna, Italy}
\\
\\
$^b${\em I.N.F.N., Sezione di Bologna, I.S.~FLAG}
\\
{\em viale B.~Pichat~6/2, 40127 Bologna, Italy}
\\
\\
$^c${\em Department of Physics and Astronomy, Bishop's University,}
\\
{\em 2600 College Street, Sherbrooke, Qu\'ebec, Canada J1M 1Z7}
}
\begin{document}
\maketitle
\begin{abstract}
We determine the complete space-time metric from the bootstrapped
Newtonian potential generated by a static spherically symmetric source in the
surrounding vacuum.
This metric contains post-Newtonian parameters which can be further
used to constrain the underlying dynamical theory and quantum state
of gravity. 
For values of the post-Newtonian parameters within experimental bounds,
the reconstructed metric appears very close to the Schwarzschild solution of
General Relativity in the whole region outside the event horizon. 
The latter is however larger in size for the same value of the mass compared
to the Schwarzschild case.
\par
\null
\par
\noindent
\textit{PACS - 04.70.Dy, 04.70.-s, 04.60.-m}
\end{abstract}
\newpage
\section{Introduction and motivation}
\setcounter{equation}{0}
\label{Sintro}
General Relativity predicts that the gravitational collapse of any compact source will generate
geodetically incomplete space-times whenever a trapping surface appears~\cite{HE}.
Moreover, eternal point-like sources are mathematically incompatible with the Einstein field
equations~\cite{geroch}.
A consistent quantum theory should fix this pathological classical picture of black hole formation, 
like quantum mechanics explains the stability of the hydrogen atom.
Whether this can be achieved by modifications of the gravitational dynamics solely at the
Planck scale or with sizeable implications for astrophysical compact objects remains open
to debate, because it is intrinsically very difficult to describe quantum states of strongly
interacting systems.
Strong interactions imply large nonlinearities, so that the space of classical solutions does 
not admit a vector basis for the canonical variables which are usually lifted to quantum operators.
Of course, this quantisation process can be introduced in a linearised version of any theory,
but it becomes questionable that one can then effectively obtain a reliable approximation for
the quantum state of what would classically be a strongly interacting configuration.
For instance, the physical relevance of the quantum theory of linear perturbations around a given
classical solution entirely relies on whether the chosen ``background'' is actually the one
realised in nature.
\par
In the Einstein theory of gravity, we know classical solutions, like the Schwarzschild metrics
for the interior of a homogenous spherical star and the exterior of any spherical source,
which cannot be obtained by perturbing the Minkowski vacuum. 
On the other hand, Deser~\cite{deser} conjectured that it should be possible to reconstruct 
the full dynamics of General Relativity from the Fierz-Pauli action in Minkowski space-time
by adding gravitational self-coupling terms consistent with diffeomorphisms invariance.
On a closer inspection, this reconstruction of the Einstein-Hilbert action does not appear
free of ambiguities since, for instance, it involves fixing the very important boundary terms in a specific
way~\cite{rubio}.
Generically, we know that any (modified) metric theory of gravity is invariant under changes of
coordinates and must therefore be covariant under diffeomorphisms.
Different choices of those boundary terms in the reconstruction proposed by Deser
would therefore lead to different modified theories of gravity.
What we do not know {\em a priori\/} is which (if any) of such theories describes the dynamics
realised in nature and what the quantum state of the Universe really is.~\footnote{We also remark that Lovelock's
theorem~\cite{lovelock} only holds in the vacuum, whereas our Universe is obviously a very different state
and so are astrophysical compact objects.}  
Moreover, any reconstruction of the dynamics starting from the Minkowski vacuum can be practically
effective only if the contribution of matter sources is perturbatively small, which introduces the further
problem of reconstructing a large astrophysical source along with the ensuing gravitational field.
Such considerations inspired a programme called
{\em bootstrapped Newtonian gravity\/}~\cite{BootN,Casadio:2020mch}, which
consists in adding gravitational self-coupling terms to a Fierz-Pauli-type of action for the static Newtonian
potential generated by an arbitrarily large matter source.
Furthermore, the coupling constants for such additional terms are allowed to vary from their
Einstein-Hilbert values in order to effectively accommodate for corrections arising 
from the underlying dynamics which, as mentioned above, we do not wish to restrict {\em a priori}.
The direct outcome of this programme is a nonlinear equation, which determines
the gravitational potential acting on test particles at rest, and which is generated by a static large source,
including pressure effects and the gravitational self-interaction to next-to-leading
order in the Newton constant.~\footnote{One could ideally iterate the process to any order,
but the equations become quickly intractable analytically.}
It is important to remark that our main aim eventually is to investigate the actual quantum state
of such systems and the resulting bootstrapped Newtonian potential must therefore be viewed as a mean-field
result depending on effective coupling constants which entail properties of such a (otherwise unknown) state.
Our approach is not meant to provide solutions of the linearised Einstein equations
(or any modifications thereof), but to describe features of the proper quantum state of gravity.
Compact objects were studied with this equation~\cite{Casadio:2019cux,Casadio:2020kbc,Casadio:2019pli}
and, at least for the simplest case of homogenous density, one can explicitly build a coherent quantum state
(for a scalar field) which reproduces the classical gravitational
potential~\cite{Casadio:2016zpl,Casadio:2017cdv}.
Interestingly, these quantum states share some of the properties~\cite{ciarfella} found in the corpuscular model
of black holes~\cite{DvaliGomez,giusti}.
\par
Accurate descriptions of the interior of matter sources, whether it is a black hole or a more regular,
yet highly compact, distribution, should be given in terms of quantum physics, possibly resulting in
an effective equation of state.
The relevant observables would eventually be represented by the radius and the mass of stable
configurations.
Instead, the exterior region of any astrophysical compact object is phenomenologically characterised by the (geodesic)
motion of test particles, including photon trajectories.
Studying these trajectories, and comparing them with those predicted by General Relativity, is more directly done
by means of a full (effective) metric tensor, rather than the bootstrapped Newtonian potential describing only forces
which act on static particles.
\par
The aim of this work is precisely to reconstruct a complete space-time metric from the bootstrapped Newtonian
potential in the vacuum outside a spherically symmetric source.
Of course, by employing an effective metric tensor we implicitly assume the effective dynamics is also invariant
under changes of coordinates, which is compatible with the underlying fundamental theory of gravity
being covariant under diffeomorphisms, although the particular metric we will find does not need to be a solution
of the Einstein equations in the vacuum.
Moreover, we will express this metric in terms of quantities which, if not directly observable, have at least an
intrinsic geometric meaning. 
In particular, we will take advantage of the spherical symmetry and employ the usual angular coordinates
on the spheres (as surfaces of symmetry of the system) of area $\mathcal A=4\,\pi\,\bar r^2$, along
with the areal radius $\bar r$.
The latter differs from the radial coordinate $r$ associated with the harmonic coordinates used to express
the potential~\cite{weinberg}, which is a source of significant technical complication.
Furthermore, starting from the potential acting on test particles at rest in a given (harmonic) reference frame
does not fix the reconstructed spherically symmetric metric uniquely.
For this reason, it will be useful to write the metric in the weak-field region in terms of post-Newtonian
parameters, which allow for a direct comparison with experimental bounds.
This procedure should, in principle, determine the entire metric in terms of the post-Newtonian parameters 
all the way into the strong coupling region, if we could solve all equations exactly.
However, the post-Newtonian expansion fails near the horizon, so that an explicit calculation will require us
to employ also a different near-horizon expansion.
Since the potential is a smooth function of $r$, so must be the metric and the relation $\bar r=\bar r(r)$.
The coefficients in the near-horizon expansion are therefore fully determined by the post-Newtonian
parameters via matching conditions in a suitable intermediate region, but analytical expressions become
rather involved very quickly.
In the present work, we shall therefore just carry out the analysis by including the first few terms in each
of the above two expansions.
\par
The main result is that the bootstrapped metric at large distance from the source approaches
the Schwarzschild form in a way that can make it compatible with bounds from Solar system tests and
other measurements of the first post-Newtonian parameters.
The bootstrapped metric is however necessarily different from the exact Schwarzschild form, and this
can be interpreted from the point of view of General Relativity as indicating the presence of an effective fluid,
filling the space around the source with a non-vanishing energy-momentum tensor which violates
the classical energy conditions.
The presence of an effective fluid in bootstrapped Newtonian gravity was already noted in
Refs.~\cite{cosmo}. 
Moreover, the near-horizon region differs from the General Relativistic prediction mostly in that
the horizon size is larger than the Schwarzschild radius for given black hole mass.
\par
The paper is organised as follows:
in Section~\ref{S:boot}, we review the derivation of the bootstrapped Newtonian potential
acting on a static test particle generated by a static spherically symmetric source.
In Section~\ref{S:harmonic}, we discuss the relation between the harmonic coordinates
used to express the potential and the more common areal radius.
This relation plays a crucial role in the reconstruction of the metric performed at
large distance from the source in Section~\ref{S:weak}, where corrections to the perihelion
precession, light deflection and time delay are also estimated.
The geometry near the horizon is studied in Section~\ref{S:horizon} by matching
with the weak-field expressions.
We conclude with comments and an outlook in Section~\ref{S:conc}.
\section{Potential in the vacuum}
\label{S:boot}
\setcounter{equation}{0}
In General Relativity (and metric theories of gravity in general), the motion of test particles is
determined by the entire metric tensor and there is no invariant notion of a gravitational potential.
However, one can still introduce a potential for specific types of motion on specific metric space-times
starting from the corresponding geodesic equation.
For example, the geodesic equation in the weak field and non-relativistic limit reduces
to the Newtonian equation of motion with the potential which solves the linearised Einstein equations
in the vacuum provided one uses harmonic coordinates.
In the following, we will reverse this argument and start from a bootstrapped Newtonian
potential obtained in harmonic coordinates in order to reconstruct a compatible metric.
\subsection{Potential for static test particles}
We consider a massive particle moving along the trajectory $x^\mu=x^\mu(\tau)$
that satisfies the geodesic equation
\begin{equation}
\label{geodesic}
    \ddot x^\mu
    +\Gamma^\mu_{\alpha\beta}\,\dot x^\alpha\,\dot x^\beta
    =0
    \ ,
\end{equation}
where dots denote derivatives with respect to the particle's proper time $\tau$ and
$\Gamma^\mu_{\alpha\beta}$ are the Christoffel symbols of the metric
$g_{\mu\nu}$.
If the space-time is static, one can choose a time coordinate $x^0$ in which the metric
reads 
\be
g_{\mu\nu}=\eta_{\mu\nu}
+
\epsilon\,h_{\mu\nu}(x^i)
\ ,
\label{eq:metric}
\ee
where $\epsilon$ is a parameter we introduce to keep track of deviations from flat 
space-time.
We can now say that the particle is (initially) at rest if $\dot x^i=0$ in this reference frame,
which implies that $\dot x^0\simeq 1$ and, as long as $|\dot x^i|\simeq\epsilon\ll 1$
(weak-field approximation), Eq.~\eqref{geodesic} to first order in $\epsilon$ reduces to
\begin{equation}
    \dot x^i
    \simeq
    \frac{1}{2}\,\epsilon\,h_{00,i}
    \ ,
\label{Fma}
\end{equation}
which yields Newton's second law for a particle in the potential $V$ if we set
\begin{equation}
\label{weak field limit}
    g_{00}
    =
    -1+\epsilon\,h_{00}
    =
    -(1+2\,V)
    \ ,
\end{equation}
and the spatial coordinates $x^i$ in Eq.~\eqref{Fma} are the analogue of 
Cartesian coordinates in Newtonian mechanics.
\par
In fact, the explicit form of the potential $V$ generated by a given source can be obtained
from the linearised Einstein equations, which then reduce to the Poisson equation for the Newtonian
potential in the \textit{de~Donder gauge}
\be
\partial^\alpha h_{\alpha\mu}
-\frac{1}{2}\,\partial_\mu h
=
0
\ ,
\label{ddgauge}
\ee
where $h\equiv \eta^{\alpha\beta}\,h_{\alpha\beta}$.
We must correspondingly assume that the coordinates $x^\mu$ in which the components of the metric 
take the form in Eq.~\eqref{eq:metric} are \textit{harmonic} coordinates satisfying
\be
\Box x^\mu
=
0
\ .
\label{boxX}
\ee
Note that for a static metric with $|h_{ij}|\ll 1$, the condition~\eqref{ddgauge} is always
satisfied.
\subsection{Bootstrapped Newtonian vacuum}
\label{SS:vacuum}
We just recalled that the interpretation of $V$ in Eq.~\eqref{weak field limit} as the gravitational
potential for massive particles at rest is consistent with the fact that, in the same approximation, the linearised
Einstein field equations reduce to the Poisson equation of Newton's theory,
\be
\triangle V
=
4\,\pi\,\gn\,\rho
\ ,
\label{poisson}
\ee
where $\rho$ is the energy density of the static source and $\triangle$ the flat space Laplacian.
The de~Donder gauge condition~\eqref{ddgauge} implemented in the derivation of Eq.~\eqref{poisson}
was thus employed explicitly also in deriving the equation for the bootstrapped Newtonian potential $V$ from
the Einstein-Hilbert action in Ref.~\cite{Casadio:2017cdv}.
For the sake of brevity, we here review a more heuristic derivation of $V=V(r)$ outside
static and spherically symmetric sources from a bootstrapped Newtonian effective
action~\cite{Casadio:2017cdv,BootN,Casadio:2019cux,Casadio:2019pli}.
\par
We start from the Newtonian Lagrangian for a source of density $\rho=\rho(r)$, to wit
\be
L_{\rm N}[V]
=
-4\,\pi
\int_0^\infty
r^2\,\d r
\left[
\frac{\left(V'\right)^2}{8\,\pi\,\gn}
+V \rho
\right]
\label{LVn}
\ee
from which Eq.~\eqref{poisson} can be derived, and stress that the radial coordinate $r$
is the one obtained from harmonic coordinates $x^i$, as we shall see more in details in
Section~\ref{S:harmonic}.
To this action several interacting terms for the field potential $V$ will be added
for the motivation, stated in the introductory section, of describing mean-field deviations
from General Relativity induced by quantum physics.
First of all, we couple $V$ to a gravitational current proportional to its own energy density,
\be
J_V
\simeq
4\frac{\d U_{\rm N}}{\d \mathcal{V}} 
=
-\frac{\left[V'(r)\right]^2}{2\,\pi\,\gn}
\ ,
\ee
where $\mathcal{V}$ is the spatial volume and $U_{\rm N}$ the Newtonian potential energy.
Moreover, we add the ``loop correction'' $J_{\rho}\simeq-2V^2$, which couples to $\rho$ and,
since the pressure gravitates and becomes relevant for large compactness,
we also add to the energy density the term~\cite{Casadio:2019cux}~\footnote{We only consider
isotropic fluids.}
\be
J_p
\simeq
-
\,\frac{\d U_p}{\d \mathcal{V}}
= 
p
\ ,
\ee
where $U_p$ is the potential energy associated with the work done by the force responsible
for the pressure.
The total Lagrangian then reads 
\be
L[V]
&\!\!=\!\!&
L_{\rm N}[V]
-4\,\pi
\int_0^\infty
r^2\,\d r
\left[
q_V\,J_V\,V
+
3\,q_p\,J_p\,V
+
q_\rho\,J_\rho\left(\rho+3\,q_p\,p\right)
\right]
\nonumber
\\
&\!\!=\!\!&
-4\,\pi
\int_0^\infty
r^2\,\d r
\left[
\frac{\left(V'\right)^2}{8\,\pi\,\gn}
\left(1-4\,q_V\,V\right)
+\left(\rho+3\,q_p\,p\right)
V
\left(1-2\,q_\rho\,V\right)
\right]
\ ,
\label{LagrV}
\ee
where the coupling constants $q_V$, $q_p$ and $q_\rho$ can be used to track the effects of the different
contributions.
As we mentioned previoulsy, different values of these couplings would correspond
to different quantum states and depend on the underlying microscopic quantum theory of gravity and matter.
For instance, the case $q_V=q_p=q_\rho=1$ reproduces the Einstein-Hilbert action at next-to-leading order
in the expansion in $\epsilon$ in Eq.~\eqref{eq:metric} and can be naturally used as a primary
reference~\cite{Casadio:2017cdv}
(see also Refs.~\cite{Casadio:2019cux,Casadio:2019pli} for more details on the role of these coupling parameters).
Eventually, their values should be fixed by experimental constraints.
Finally, the field equation for $V$ reads
\be
\triangle V
=
4\,\pi\,\gn
\frac{1-4\,q_\rho\,V}{1-4\,q_V\,V}
\left(\rho+3\,q_p\,p\right)
+
\frac{2\,q_V\left(V'\right)^2}
{1-4\,q_V\,V}
\ ,
\label{EOMV}
\ee
which must be solved along with the conservation equation $p' = -V'\left(\rho+p\right)$. 
\par
In vacuum, where $\rho=p=0$, Eq.~\eqref{EOMV} simplifies to
\be
\triangle V
=
\frac{2\,q_V \left(V'\right)^2}{1-4\,q_V\,V}
\ ,
\label{EOMV0}
\ee
which allows for absorbing the coupling constant $V\to \tilde V=q_V\,V$. 
The exact solution was found in Ref.~\cite{BootN} and reads
\be
V_{0}
=
\frac{1}{4\,q_V}
\left[
1-\left(1+\frac{6\,q_V\,\gn\,M}{r}\right)^{2/3}
\right]
\ .
\label{sol0}
\ee
The asymptotic expansion away from the source yields
\be
V_{0}
\simeq
-\frac{\gn\,M}{r}
+q_V\,\frac{\gn^2\,M^2}{r^2}
-q_V^2\,\frac{8\,\gn^3\,M^3}{3\,r^3}
\ ,
\label{Vasy}
\ee
so that the Newtonian behaviour is always recovered and the post-Newtonian terms
are seen to depend on the coupling $q_V$.
The value of $q_V$ can be constrained by experimental bounds once
we compute trajectories to compare with. 
\section{Harmonic and areal coordinates for static spherical systems}
\label{S:harmonic}
\setcounter{equation}{0}
The argument leading to the potential~\eqref{sol0} starting from a general metric
involves several approximations, which makes it impossible to determine the
starting metric uniquely.
In order to reconstruct a metric compatible with Eq.~\eqref{sol0}, we will therefore have
to supply further conditions.
Before we get to that point, however, we need to discuss in details the relation
between the radial coordinate $r$ used to express the potential in the previous
section and the areal coordinate $\bar r$ usually employed to write the general static
spherically symmetric metric as
\be
\label{standard form}
\d s^2
=
-\bar B\, \d \bar t^2
+\bar A\,\d \bar r^2
+\bar r^2\,\d \Omega^2
\ ,
\ee
where $\bar A=\bar A(\bar r)$, $\bar B=\bar B(\bar r)$, and
$\d \Omega^2=\d \theta^2+\sin^2\theta\,\d \phi^2$ is the usual solid angle
on the unit sphere, with $0\le\theta\le\pi$ and $0\le\phi<2\,\pi$.
\par
Cartesian coordinates $x^i=(x,y,z)$ in flat space satisfy Eq.~\eqref{boxX}.
This condition can be extended to general space-times by defining harmonic coordinates
$x^\mu=(t,x,y,z)=(t,\bm x)$ such that 
\be
\Box x^\mu
=
g^{\alpha\beta}\,\Gamma_{\alpha\beta}^\mu
=
0
\ ,
\label{eq:boxx}
\ee
which coincides with the de~Donder gauge condition~\eqref{ddgauge}.
In particular, we are interested in spherically symmetric space-times with a metric
of the form~\eqref{standard form} and we therefore find it convenient
to employ polar coordinates associated to the harmonic ones by
\be
    \label{from standard to harmonic}
    x=r(\bar r)\,\sin\theta\,\cos\phi
    \ ,
    \quad
    y=r(\bar r)\,\sin\theta\,\sin\phi
    \ ,
    \quad
    z=r(\bar r)\,\cos\theta
    \ ,
\ee
where we assume that the ``harmonic''~\footnote{Polar coordinates do not satisfy Eq.~\eqref{eq:boxx}
even in Minkowski space-time, but we shall refer to $r$ as the ``harmonic'' radial coordinate for the sake of brevity.}
$r$ is an invertible smooth function of the areal coordinate $\bar r$.
A straightforward calculation of Eq.~\eqref{eq:boxx} reveals that the function $r=r(\bar r)$
must satisfy~\cite{weinberg}
\be
\label{equation for harmonic coordinate}
\frac{\d}{\d\bar r}
\left(
\bar r^2\,\sqrt{\frac{\bar B}{\bar A}}\,\frac{\d r}{\d\bar r}
\right)
=2\,\sqrt{\bar A\,\bar B}\,r
\ .
\ee
Expressing the metric~\eqref{standard form} in terms of the the rotationally invariant forms
$\d \bm x^2=\d r^2+r^2\,\d\Omega^2$ and $(\bm x\cdot \d \bm x)^2=r^2\,\d r^2$, we deduce that the line element
in harmonic coordinates reads
\be
\label{harmonic form}
\d s^2
=
-B\,\d t^2
+
\frac{\bar r^2}{r^2}\,\d \bm x^2
+\left[
A\left(\frac{\d\bar r}{\d r}\right)^2
-\frac{\bar r^2}{r^2}
\right]
\frac{(\bm x\cdot \d \bm x)^2}{r^2}
\ ,
\ee
where $\d t=\d\bar t$, $\bar r=\bar r(r)$, $A=\bar A(\bar r(r))$ and $B=\bar B(\bar r(r))$.
\par
\begin{figure}[t]
\centering
\includegraphics[width=8cm]{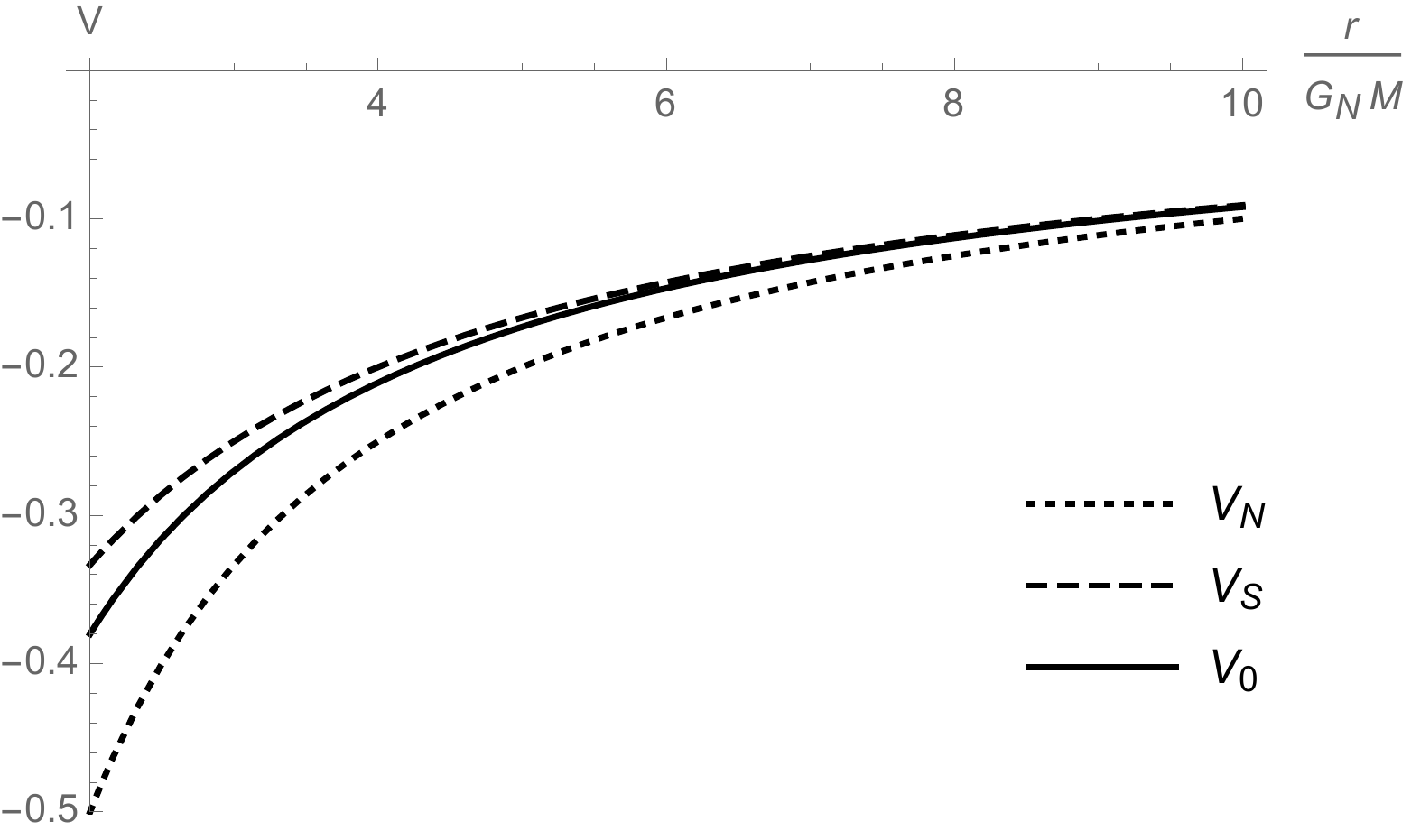}
\caption{Comparison between the Newtonian $V_{\rm N}$, Schwarzschild $V_{\rm S}$
and bootstrapped Newtonian $V_{0}$ (with $q_V=1$).}
\label{f:potentials}
\end{figure}

The unique Schwarzschild solution of the Einstein field equations in the vacuum outside a spherical
source~\footnote{Birkhoff's theorem ensures that uniqueness follows from spherical symmetry.
In more general cases, other vacuum solutions can be obtained from the linearised solutions~\cite{xanto}.}
is given by
\be
\bar B_{\rm S}
=
\frac{1}{\bar A_{\rm S}}
=
1
-\frac{\Rh}{\bar r}
\ ,
\label{schw}
\ee 
where
\be
\Rh
=
2\,\gn\,M
\label{Rh}
\ee
is the gravitational radius.
By solving Eq.~\eqref{equation for harmonic coordinate}, one finds that the harmonic
radial coordinate for the Schwarzschild metric is simply given by 
\be
r
=
\bar r
-
\frac{\Rh}{2}
\equiv
\bar r
-r_{\rm S}
\ ,
\label{rSbr}
\ee
which leads to the potential for the Schwarzschild metric in harmonic coordinates
\be
V_{\rm S}
=
\frac{1}{2}
\left[B_{\rm S}-1\right]
=
-
\frac{\gn\,M}{r}
\left(1
+\frac{\gn\,M}{r}
\right)^{-1}
\ .
\label{Vs}
\ee
By comparing with the expansion of $V_0$ in Eq.~\eqref{Vasy}, we then see that the unique
prediction of General Relativity is recovered to first order in $q_V$ if $q_V=1$
(see Fig.~\ref{f:potentials}).
\par
We can now replace $V_{\rm S}$ with the potential $V_0$ in Eq.~\eqref{sol0}, that is
\be
B
=
1+2\,V_0
\ ,
\label{Bbn}
\ee
and start to reconstruct the bootstrapped metric in the areal coordinate $\bar r$.
In particular, we notice that the metric coefficient $\bar B$ is fully determined
by the potential $V_0$ and the relation $r=r(\bar r)$.
Moreover, the Schwarzschild metric has the important property that
$\bar A_{\rm S}\,\bar B_{\rm S}=1$, 
which is related with the vanishing of the light-like component of the Ricci tensor,
$R_{\mu\nu}\,k^\mu\,k^\nu=0$ for any $k_\mu\,k^\mu=0$~\cite{jacobson},
and the validity of the Equivalence Principle.
Using $\bar C\equiv \bar A\,\bar B$, it is also convenient to rewrite
Eq.~\eqref{equation for harmonic coordinate} as
\be
\label{equation for harmonic coordinate second form}
\bar r\,r''
+
\left(
2
-\frac{\bar r\,{\bar C'}}{2\,\bar C}
+\bar r\,\frac{\bar B'}{\bar B}
\right)
r'
=
2\,\frac{\bar C\,r}{\bar B\,\bar r}
\ ,
\ee
where a prime denotes the derivative with respect to $\bar r$.
This equation determines the relation between $\bar A$ and $\bar r$, but one equation
is not sufficient to determine both $r=r(\bar r)$ and $\bar A=\bar A(\bar r)$ given $B=B(r)$,
and we will have to resort to further conditions.
\section{Effective space-time picture: weak field}
\label{S:weak}
\setcounter{equation}{0}
We first analyse the region far from the source by Taylor expanding the metric coefficients
and $r=r(\bar r)$ in powers of the dimensionless ratio $\Rh/\bar r\sim M/\bar r$, that is
\be
\begin{split}
&
\bar A
=
1+\sum_{k=1} a_k \pqty{\frac{\Rh}{\bar r}}^k
\ ,
\\
&
\bar B
=
1+\sum_{k=1} b_k \pqty{\frac{\Rh}{\bar r}}^k
\end{split}
\label{expWg}
\ee
and
\be
\frac{r}{\bar r}
=
1+\sum_{k=1} \sigma_k \pqty{\frac{\Rh}{\bar r}}^k
\ .
\label{expWr}
\ee
We also introduce
\be
\bar C
=
1+ \sum_{k=1} c_k \pqty{\frac{\Rh}{\bar r}}^k
\ ,
\ee
in which the coefficients $a_k$'s are fully determined by the $c_k$'s and
$b_k$'s since $\bar C=\bar A\,\bar B$.
The above expressions for $\bar C$, $\bar B$ and $r/\bar r$ solve
Eq.~\eqref{equation for harmonic coordinate second form}
[equivalently, Eq.~\eqref{equation for harmonic coordinate}] at zero order
in $\Rh/\bar r$ and ensure asymptotic flatness for $r\sim\bar r\to\infty$.
In the following, we will solve Eq.~\eqref{equation for harmonic coordinate second form} 
in order to determine the metric up to third order in $\Rh/\bar r$.
\par
At first and second order in $\Rh/\bar r$ we obtain
\be
\label{sigma coefficients}
\begin{split}
&
\sigma_1
=
\frac{b_1}{2}
-\frac{3}{4}\,c_1
\\
&
\sigma_2
=
\frac{c_1}{4}\left(
2\,c_1-b_1
\right)
-
\frac{c_2}{2}
\ ,
\end{split}
\ee
and the third-order equation yields
\begin{equation}
\label{c3 coefficient}
c_3
=
\frac{5}{2}\,c_1\,c_2
-\frac{1}{2}\,b_1^2\,c_1
-b_1\,c_2
+b_2\,c_1
+\frac{5}{4}\,b_1\,c_1^2
-2\,b_3
-\frac{3}{2}\,c_1^3
\ .
\end{equation}
We can now fix the coefficients $b_k$ to match Eq.~\eqref{Bbn}, that is
\be
\bar B
\simeq
1
-\frac{\Rh}{r(\bar r)}
+\frac{q_V\,\Rh^2}{2\,[r(\bar r)]^2}
-\frac{2\,q_V^2\,\Rh^3}{3\,[r(\bar r)]^3}
\ ,
\ee
which yields $b_1=-1$ and
\be
\begin{split}
&
b_2
=
\frac{q_V}{2}
-\frac{3}{4}\,c_1
-\frac{1}{2}
\\
&
b_3
=
\frac{q_V}{4}
\left(
2+3\,c_1
\right)
-\frac{2}{3}\,q_V^2
-\frac{c_1}{16}
\left(8+c_1\right)
-\frac{c_2}{2}
-\frac{1}{4}
\ .
\end{split}
\ee
Upon replacing the above expressions in the expansion of $\bar A$, we obtain
\be
\begin{split}
&
a_1
=
1+c_1
\\
&
a_2
=
\frac{3}{2}
-\frac{q_V}{2}
+\frac{7}{4}\,c_1
+c_2
\\ 
&
a_3
=
\frac{11}{4}
+
\left(2\,q_V
-
\frac{5}{2}
-\frac{9}{4}\,c_1
\right)
q_V
+
\frac{7}{2}\,
(c_1+c_2)
+
\frac{c_1}{2}
\left(
5\,c_2
-
\frac{17}{8}\,c_1
-
3\,c_1^2
\right)
\ .
\end{split}
\ee
\par
In order to uniquely fix all of the coefficients in the above expansions from physical considerations,
it is useful to introduce the Eddington-Robertson parameterised post-Newtonian (PPN) formalism,
in which the metric reads~\cite{weinberg}
\be
\d s^2
\simeq
-\left[
1-\alpha\,\frac{\Rh}{\bar r}
+(\beta-\alpha\,\gamma)\,\frac{\Rh^2}{2\,\bar r^2}
+(\zeta-1)\,\frac{\Rh^3}{\bar r^3}
\right]
\d \bar t^2
+
\left[
1+\gamma\,\frac{\Rh}{\bar r}
+\xi\,\frac{\Rh^2}{\bar r^2}
\right]
\d \bar r^2
+\bar r^2\,\d\Omega^2
\ ,
\ee
where one can set $\alpha=1$ by the definition of the gravitational radius~\eqref{Rh}.
This is in agreement with $b_1=-\alpha=-1$ and allows us to identify
the first order PPN parameters 
\be
c_1=\gamma-1
\quad
{\rm and}
\quad
q_V=\beta+\frac{\gamma-1}{2}
\ .
\label{qVppn}
\ee
Finally, we obtain
\be
\bar B
&\!\!\simeq\!\!&
1
-\frac{\Rh}{\bar r}
+
\left(\beta-\gamma\right)
\frac{\Rh^2}{2\,\bar r^2}
\nonumber
\\
&&
+
\left[
7+4\,\beta\,(5+\gamma)-32\,\beta^2-\gamma\,(26-7\,\gamma)-24\,c_2
\right]
\frac{\Rh^3}{48\,\bar r^3}
\ee
and
\be
\bar A
&\!\!\simeq\!\!&
1
+
\gamma\,\frac{\Rh}{\bar r}
-
\left(
\beta-3\,\gamma-2\,c_2
\right)
\frac{\Rh^2}{2\,\bar r^2}
\nonumber
\\
&&
+
\left[
5
+
32\,\beta^2
-
4\,\beta\,(9+\gamma)
+
3\,\gamma\,(6+15\,\gamma-8\,\gamma^2)
+
8\,c_2\,(2+5\,\gamma)
\right]
\frac{\Rh^3}{16\,\bar r^3}
\ee
so that
\be
\bar C
&\!\!\simeq\!\!&
1
+
(\gamma -1)\,\frac{\Rh}{\bar r}
+
c_2\,\frac{\Rh^2}{\bar r^2}
\\
&&
+
\left[
11
+
32\,\beta^2
-
8\,\beta\,(4-\gamma)
-
\gamma\,(22-59\,\gamma-36\,\gamma^2)
-
12\,c_2\,(1-5\,\gamma)
\right]
\frac{\Rh^3}{24\, \bar r^3}
\ .
\label{Cppn}
\ee
The harmonic radius is also given by
\be
r
\simeq
\bar r
+\frac{1-3\,\gamma}{4}\,
\Rh
+
\left(
1-3\,\gamma
+2\,\gamma^2
-2\,c_2
\right)
\frac{\Rh^2}{4\,\bar r}
\ .
\ee
\par
Experimental data strongly constrain $\abs{\gamma-1}\simeq \abs{\beta-1}\ll 1$.
Upon assuming $\beta=\gamma=1$, that is $c_1=0$ and $q_V=1$, we find that the
bootstrapped metric which describes the minimum deviation from the Schwarzschild form
is given by
\be
\bar B
&\!\!\simeq\!\!&
1
-\frac{2\,\gn\,M}{\bar r}
-
2\left(5+6\,c_2\right)
\frac{\gn^3\,M^3}{3\,\bar r^3}
\nonumber
\\
&\!\!\simeq\!\!&
B_{\rm S}(\bar r)
-
2\left(6\,\xi-1\right)
\frac{\gn^3\,M^3}{3\,\bar r^3}
\label{Bc2}
\\
\bar A
&\!\!\simeq\!\!&
1
+
\frac{2\,\gn\,M}{\bar r}
+
4\left(
1+c_2
\right)
\frac{\gn^2\,M^2}{\bar r^2}
+
2\,\left(9+14\,c_2\right)
\frac{\gn^3\,M^3}{\bar r^3}
\nonumber
\\
&\!\!\simeq\!\!&
A_{\rm S}(\bar r)
+
(\xi-1)\,
\frac{\gn^2\,M^2}{\bar r^2}
+
2\left(14\,\xi-9\right)
\frac{\gn^3\,M^3}{\bar r^3}
\ ,
\label{Ac2}
\ee
and
\be
r
\simeq
\bar r
-
\gn\,M
-
2\,(\xi-1)\,\frac{\gn^2\,M^2}{\bar r}
\ .
\label{rwmin}
\ee
For completeness, we also display the Ricci scalar obtained
from the above metric coefficients to next-to-leading order in the $\Rh/r$ expansion,
\be
\bar R
&\!\!\simeq\!\!&
\left(
1-4\,c_2-5\,\gamma+4\,\gamma^2
\right)
\frac{\Rh^2}{2\,r^4}
\nonumber
\\
&&
-
\left(
9
+16\,c_2
-40\,\beta
+32\,\beta^2
+14\,\gamma
+16\,c_2\,\gamma
+16\,\beta\,\gamma
+5\,\gamma^2
-16\,\gamma^3
\right)
\frac{\Rh^3}{8\,r^5}
\nonumber
\\
&\!\!\simeq\!\!&
-2\,c_2
\frac{\Rh^2}{r^4}
-
\left(
5+8\,c_2
\right)
\frac{\Rh^3}{2\,r^5}
\ ,
\ee
where we set $\beta=\gamma=1$ in the second expression.
Clearly, the above expression of $\bar R$ shows that the effective metric increasingly differs
from Schwarzschild's $\bar R_{\rm S}=0$ as one goes closer to the source.
\par
In the above, the second order PPN parameters are both determined by the one
parameter $c_2$ as
\be
\xi
=
1+c_2
\ ,
\quad
{\rm and}
\quad
\zeta
=
1
-\frac{5+6\,c_2}{12}
=
\frac{13-6\,\xi}{12}
\ ,
\label{2PN}
\ee
so that the combination $\xi=\zeta=1$ corresponding to the PPN expansion of the
Schwarzschild metric is not allowed.
We can see that the new contribution to $\bar A$ at second order in $\Rh/\bar r$ only vanishes
for $c_2=0$, but higher-order corrections then cannot be eliminated.
Correspondingly, for $\beta=\gamma=1$, we have 
\be
\bar C
\simeq
1
+
(\xi-1)\,\frac{\Rh^2}{\bar r^2}
+
(12\,\xi-7)\,
\frac{\Rh^3}{6\, \bar r^3}
\ ,
\label{bCw}
\ee
and the Schwarzschild case $\bar C=1$ cannot be reproduced.
In the following, we shall analyse the effects of these second order terms in
Eqs.~\eqref{Bc2} and~\eqref{Ac2}.
\subsection{Effective energy-momentum tensor}
\label{SS:emt}
Since the effective metric with components~\eqref{Bc2} and~\eqref{Ac2} 
differs from the Schwarzschild geometry, the space-time must contain a non-vanishing
effective spherically symmetric energy-momentum tensor 
\be
T_{\mu\nu}^{\rm eff}
=
\rho^{\rm eff}\,u_\mu\, u_\nu
+
p_r^{\rm eff}\,r_\mu\, r_\nu
+
p_t^{\rm eff}\,\theta_\mu\, \theta_\nu
+
p_t^{\rm eff}\,\phi_\mu\, \phi_\nu
\ ,
\ee
where $\rho^{\rm eff}=\rho^{\rm eff}(\bar r)$, $p_r^{\rm eff}=p_r^{\rm eff}(\bar r)$
and $p_t^{\rm eff}=p_t^{\rm eff}(\bar r)$ are respectively
the energy density, the radial pressure and the surface tension of the static effective fluid.
In the coordinates $\bar x^\mu=(\bar t,\bar r,\theta,\phi)$ of Eq.~\eqref{standard form},
we also have the tetrad components
\be
u^\mu
=
\frac{\delta^\mu_0}{\bar B^{1/2}}
\ ,
\qquad
r^\mu
=
\frac{\delta^\mu_{1}}{\bar A^{1/2}}
\ ,
\qquad
\theta^\mu
=
\frac{\delta^\mu_{3}}{\bar r}
\ ,
\qquad
\phi^\mu
=
\frac{\delta^\mu_{4}}{\bar r\,\sin\theta}
\ .
\ee
We can compute the density and pressures from the Einstein tensor,
\be
\rho^{\rm eff}
&\!\!=\!\!&
T^{\rm eff}_{\mu\nu}\,u^\mu\, u^\nu
=
\frac{G_{00}}{8\,\pi\,\gn\,\bar B}
=
\frac{(\bar A-1)\, \bar A+\bar r \,\bar A'}{8\, \pi\,  \gn\,\bar r^2\, \bar A^2}
\nonumber
\\
p^{\rm eff}_r
&\!\!=\!\!&
T^{\rm eff}_{\mu\nu}\,r^\mu\,r^\nu
=
\frac{G_{11}}{8\,\pi\,\gn\,\bar A}
=
\frac{\bar B-\bar C+\bar r\, \bar B'}{8\, \pi\, \gn\, \bar r^2\, \bar C}
\\
p^{\rm eff}_t
&\!\!=\!\!&
T^{\rm eff}_{\mu\nu}\,\theta^\mu\,\theta^\nu
=
\frac{G_{22}}{8\,\pi\,\gn\,\bar r^2}
=
\frac{2\,\bar C \left(2\, \bar B'+\bar r\, \bar B''\right)-\left(2\, \bar B+\bar r\, \bar B'\right) \bar C'}
{32\, \pi\, \gn\, \bar r\, \bar C^2}
\ ,
\nonumber
\ee
where a prime denotes differentiation with respect to $\bar r$.
The above expressions of course vanish for the Schwarzschild metric, whereas we
obtain~\footnote{The general expressions in terms of Eddington-Robertson
parameters is given in Appendix~\ref{A:emt}.}
\be
\rho^{\rm eff}
&\!\!\simeq\!\!&
\frac{\gn\,M^2}{2\,\pi\,\bar r^4}
\left[
1-\xi
+
(1-6\,\xi)\,
\frac{\gn\,M}{\bar r}
\right]
\\
p_{r}^{\rm eff}
&\!\!\simeq\!\!&
(1-\xi)\,\frac{\gn\,M^2}{2\,\pi\,\bar r^4}
\left(
1
+
\frac{2\,\gn\,M}{\bar r}
\right)
\nonumber
\\
&\!\!\simeq\!\!&
-p_{t}^{\rm{eff}}
\ .
\ee
For $\xi=1$ (that is $c_2=0$), the pressure and tension vanish, at this order of approximation,
but one is still left with a negative energy density.
\subsubsection{Energy conditions}
One can now check if the effective source satisfies (some of) the
energy conditions.
Since $p_r\simeq -p_t$ the effective fluid is in general anisotropic. 
In particular, for anisotropic fluids, the null energy condition is implied by all
other energy conditions and requires
\be
&&
0
\le
\rho+p_r
=
\frac{\bar B\,\bar C'}{8\, \pi\,  \gn\,\bar r\,\bar C^2}
\label{nec1}
\\
&&
0
\le
\rho+p_t
=
\frac{2\,\bar r^2\, \bar C\, \bar B''-\bar r^2 \,\bar B'\, \bar C'+\bar B \left(2\,\bar r\, \bar C'-4\, \bar C\right)+4\, \bar C^2}
{32 \,\pi\, \gn\,\bar r^2\,\bar C^2}
\ ,
\label{nec2}
\ee
where primes again denote differentiation with respect to $\bar r$.
\par
For $\beta=\gamma=1$, we have 
\be
&&
\rho+p_r
\simeq
\frac{\gn\,M^2}{\pi\,\bar  r^4}
\left[
1-\xi
+
(3-8\,\xi)\,\frac{ \gn\, M}{2\,\bar r}
\right]
\label{nec11}
\\
&&
\rho+p_t
\simeq
-
(1+4\,\xi)\, 
\frac{\gn^2\,M^3}{2\, \pi\,\bar  r^5}
\ ,
\label{nec22}
\ee
and, in order to enforce the above conditions~\eqref{nec1} and~\eqref{nec2}
for $\bar r\gg \Rh$, we would then need $\xi<-1/4$ (that is, $c_2<-5/4$). 
The case $\xi=1$ (or $c_2=0$) of minimal deviation from the Schwarzschild metric 
necessarily violates the classical energy conditions.
In principle, this conclusion is in line with the original idea that
the effective metric should incorporate corrections stemming from quantum
physics.
The fact that the effective energy-momentum tensor does not vanish at large
distance from the source means that quantum effects associated with a localised
source will affect the space-time even at much larger scales.
\subsubsection{Misner-Sharp-Hernandez mass}
It is also interesting to cast the above result in terms of the Misner-Sharp-Hernandez
mass~\cite{MSH-1,MSH-2,MSH-3}
\be
m(\bar r)
=
\frac{\bar r}{2\,\gn}
\left(
1-\frac{1}{A(\bar r)}
\right)
\ ,
\ee
which is known to play an important role in the study of the viability of quantum and
quantum-corrected black hole solutions (see {\em e.g.}~\cite{VF1,VF2} and references
therein).\footnote{It is also worth mentioning that the Misner-Sharp-Hernandez 
has a role in determining the location of horizons for static spherically symmetric spacetimes, thus
providing a straightforward method for the characterization of the causal structure of such spaces
(see {\em e.g.}~Ref.~\cite{MSH-3}).}
For $\beta=\gamma=1$, we find
\be
m(\bar r)
\simeq
M
\left[
1
+
(\xi-1)\,\frac{2\,\gn\, M}{\bar r}
+
(6\,\xi-1)\,\frac{\gn^2\,M^2}{\bar r^2}
\right]
\ ,
\ee
which equals
\be
m(\bar r)
=
M_{\rm s}
+
4\,\pi
\int_{\bar r_{\rm s}}^{\bar r}
\rho^{\rm eff}(x)\,x^2\,\d x
\ ,
\ee
where $\bar r_{\rm s}\gg \Rh$ is the areal radius of the source of mass $M_{\rm s}=m(\bar r_{\rm s})$.
For $\xi\ge 1$ (or $c_2\ge 0$), one therefore finds that the asymptotic ADM~\cite{adm} mass
$m(\bar r\to \infty)=M<M_{\rm s}$ (the effective negative energy density screens gravity),
whereas for $\xi<1$ (or $c_2<0$) we have $M>M_{\rm s}$
(the positive effective energy density causes an anti-screening effect~\cite{Bonanno:2019ilz}).
\subsection{Geodesics}
\label{ss:geodesics}
Geodesics $\bar x^\mu=\bar x^\mu(\lambda)$ in a metric of the form in Eq.~\eqref{standard form} can be obtained
from the Lagrangian
\be
2\,L
=
\bar B\,\dot {\bar t}^2
-\bar A\,\dot {\bar r}^2
-\bar r^2
\left(
\dot\theta^2
+\sin^2\theta\,\dot\phi^2
\right)
=
k
\ ,
\label{Lgeo}
\ee
where a dot denotes differentiation with respect to $\lambda$.
The constant $k=1$ and $\lambda=\tau$ is the proper time for massive particles, whereas $k=0$ and
$\lambda$ is an affine parameter for light signals.
Staticity and spherical symmetry ensure the existence of the usual integrals of motion,
namely
\be
\mathcal{E}
=
\frac{\partial L}{\partial\dot t}
=
\bar B\,
\dot{\bar t}
\label{scE}
\ee
and
\be
\mathcal J
=
-\frac{\partial L}{\partial\dot\phi}
=
\bar r^2\,\dot\phi
\ ,
\ee
which is proportional to the angular momentum around the axis that defines the angle $\phi$
having chosen the trajectory to lie on the plane $\theta=\pi/2$.
\par
We are now left with just the equations of motion for $\phi=\phi(\tau)$ and $r=r(\tau)$,
for which it is easier to use the mass-shell condition~\eqref{Lgeo}, which we write as
\be
\dot{\bar r}^2
+
\mathcal V_{\rm eff}
=
\frac{\mathcal E^2}{\bar C}
\ ,
\label{GeoRa}
\ee
where the effective potential 
\be
\mathcal V_{\rm eff}
=
\frac{1}{\bar A}
\left(k+\frac{\mathcal J^2}{\bar r^2}\right)
\ .
\label{VeffR}
\ee
An interesting feature is that $\bar C=\bar A\,\bar B\not= 1$ in general, see Eq~\eqref{Cppn}, and one
therefore expects an energy-dependent term in the acceleration experienced by a particle,
in apparent violation of the equivalence principle~\cite{gasperini}, as predicted by some quantum models
of gravity~\cite{Bjerrum-Bohr:2015vda}.  
\par
For the purpose of studying orbits with $\mathcal J\not=0$, it is more useful
to parameterise the trajectories with the angle $\phi$, and therefore solve
\be
\left(
\frac{\d\bar r}{\d \phi}
\right)^2
=
\left(
\frac{\dot{\bar r}}{\dot\phi}
\right)^2
\ .
\ee
We next analyse massive ($k=1$) and massless ($k=0$) cases separately.
\subsubsection{Perihelion precession}
The precession of almost Newtonian orbits of planets and stars ($k=1$)
with {\em semilatus rectum} $\ell$ and eccentricity $\varepsilon$ can be easily expressed
in terms of the PPN parameters.
In particular, at first order in $\Rh/\ell$, one finds~\cite{weinberg}
\be
\Delta\phi^{(1)}
=
2\,\pi\,
(2-\beta+2\,\gamma)\,
\frac{\gn\,M}{\ell}
\ ,
\ee
which reproduces the General Relativistic result 
\be
\Delta\phi^{(1)}_{\rm S}
=
6\,\pi\,
\frac{\gn\,M}{\ell}
\ee
for $\beta=\gamma=1$ of the Schwarzschild metric.
The second order correction depends on both $\xi$ and $\zeta$ and,
for $\beta=\gamma=1$, is given by
\be
\Delta\,\phi^{(2)}
&\!\!=\!\!&
\pi
\left[
(41+10\,\xi-24\,\zeta)
+
(16\,\xi-13)\,\frac{\varepsilon^2}{2}
\right]
\frac{\gn^2\,M^2}{\ell^2}
\nonumber
\\
&\!\!\simeq\!\!&
\pi
\left[
(37+22\,c_2)
+
(3+16\,c_2)\,\frac{\varepsilon^2}{2}
\right]
\frac{\gn^2\,M^2}{\ell^2}
\nonumber
\\
&\!\!\simeq\!\!&
\Delta\,\phi^{(2)}_{\rm S}
+
2\,\pi
\left[
11\,\xi-7
+
4\,(\xi-1)\,\varepsilon^2
\right]
\frac{\gn^2\,M^2}{\ell^2}
\ ,
\ee
where we used Eq.~\eqref{2PN}, and the General Relativistic result $\Delta\,\phi^{(2)}_{\rm S}$
corresponds to $\xi=\zeta=1$.
We see that, in the minimal case with $c_2=0$, we have $\xi=1$ but $\zeta\not=1$, and a correction remains
which is independent of the eccentricity.
Binary systems could therefore be employed in order to test the effective bootstrapped Newtonian
metric at the second PPN order.
\subsubsection{Light deflection}
For light signals ($k=0$), one can likewise express the weak lensing angle for a trajectory
reaching the minimum areal radius $\bar r_0$ from infinity in terms of the PPN parameters.
At first order in $\Rh/\bar r_0$, we have~\cite{weinberg}
\be
\Delta\phi^{(1)}
=
(1+\gamma)\,
\frac{2\,\gn\,M}{\bar r_0}
\ ,
\ee
which reproduces the result from the Schwarzschild geometry for $\gamma=1$ by construction.
The second order correction for $\beta=\gamma=1$, however, only depends on $\xi$ and is given by
\be
\Delta\phi^{(2)}
&\!\!=\!\!&
\left[
\left(\frac{11}{4}+\xi
\right)
\pi
-4
\right]
\frac{\gn^2\, M^2}{\bar r_0^2}
\nonumber
\\
&\!\!\simeq\!\!&
\left[
(15+4\,c_2)\,\pi-16
\right]
\frac{\gn^2\, M^2}{4\,\bar r_0^2}
\nonumber
\\
&\!\!\simeq\!\!&
\Delta\,\phi^{(2)}_{\rm S}
+
(\xi-1)\,\pi\,\frac{\gn^2\, M^2}{\bar r_0^2}
\ ,
\ee
which equals the General Relativistic result in the minimal case $\xi=1$ (or $c_2=0$).
This shows that light is not significantly affected and weak gravitational lensing 
cannot be efficiently used to test the bootstrapped Newtonian metric.
\subsubsection{Time delay}
The radial equation~\eqref{GeoRa} for $\beta=\gamma=1$ reads
\be
\dot{\bar r}^2
&\!\!\simeq\!\!&
-k \left\{
1-\frac{2 \,\gn\, M}{\bar r}
\left[1
+\frac{2\,c_2\,\gn\, M}{\bar r}
+(5+6\,c_2)\,\frac{\gn^2\,M^2}{\bar r^2}
\right]
\right\}
-
\frac{\mathcal J^2}{\bar r^2}
\left(1-\frac{2\,\gn\, M}{\bar r}\right)
\nonumber
\\
&&
+
\mathcal E^2
\left\{
1
-
\frac{4\,\gn^2\, M^2}{\bar r^2}
\left[
c_2
+
(5+12\,c_2)\,\frac{\gn\,M}{3\,\bar r}
\right]
\right\}
\ ,
\label{GeoR}
\ee
which, even for the minimal deviation with $c_2=0$, contains an additional term 
proportional to $\mathcal E^2$.
This terms will give rise to an additional acceleration 
\be
\ddot{\bar r}
\sim
\frac{\gn^3\,M^3}{\bar r^4}\,
\mathcal E^2
\ ,
\label{accE}
\ee
which will affect the time of flight of both massive and light signals compared to 
the General Relativistic expectation.
\par
Let us consider, in particular, a trajectory with $\mathcal J=0$ between $\bar r_1=\bar r(\lambda_1)$
and $\bar r_2=\bar r(\lambda_2)>\bar r_1$.
Eq.~\eqref{GeoRa} with $c_2=0$ then reads
\be
\dot{\bar r}^2
\simeq
-k
\left[
1-\frac{2 \,\gn\, M}{\bar r}
\left(1
+\frac{5\,\gn^2\,M^2}{\bar r^2}
\right)
\right]
+
\mathcal E^2
\left(
1
-
\frac{20\,\gn^3\,M^3}{3\,\bar r^3}
\right)
\ ,
\ee
For light signals, since $k=0$,
\be
\left(
1
+
\frac{10\,\gn^3\,M^3}{3\,\bar r^3}
\right)
\dot{\bar r}
\simeq
\mathcal E
\ ,
\ee
which yields
\be
\lambda_2-\lambda_1
&\!\!\simeq\!\!&
\frac{\bar r_2-\bar r_1}
{\mathcal E}
\left[
1
+
\frac{5\,\gn^3\,M^3}{3\,\bar r_1^2\,\bar r_2^2}
\left(\bar r_1+\bar r_2\right)
\right]
\nonumber
\\
&\!\!\equiv\!\!&
\Delta\lambda
\left(1+\frac{\delta\lambda}{\Delta\lambda}\right)
\ .
\ee
The expected relative time delay $\delta\lambda/\Delta\lambda$ for light signals is therefore
independent of  $\mathcal E$.
\section{Effective space-time picture: near horizon}
\label{S:horizon}
\setcounter{equation}{0}
The task of reconstructing a metric from the potential~\eqref{sol0} is more challenging
near the horizon, as we have far less experimental constraints to rely upon.
Moreover, we need to first discuss how the horizon would be determined by the potential
in harmonic coordinates.
For the Schwarzschild solution~\eqref{schw}, the horizon areal radius is given by $\bar r=\Rh$,
which corresponds to the harmonic radius $r_{\rm S}=\Rh/2=\gn\,M$, according to Eq.~\eqref{rSbr}.
The potential~\eqref{Vs} then takes the value
\be
V_{\rm S}(r_{\rm S})=-1/2
\ ,
\ee
in agreement with the Newtonian concept of escape velocity being equal to the
speed of light.
\par
In Refs.~\cite{BootN,Casadio:2019cux,Casadio:2020kbc,ciarfella}, we relied on 
this result and likewise defined the horizon as the radius where the escape velocity
equals the speed of light for the bootstrapped Newtonian potential, that is
\be
2\,V_0(\rh)
=
-1
\ ,
\ee
which yields
\be
\rh
=
\frac{6\,q_V\,\gn\,M}{(1+2\,q_V)^{3/2}-1}
\ ,
\label{rhN}
\ee
provided $q_V>0$.
Note also that 
\be
\lim_{q_V\to 0}\rh
=
\Rh
\ ,
\ee
which is twice the Schwarzschild value $r_{\rm S}=\Rh/2$.
Considering Eq.~\eqref{qVppn} and the constraints on the PPN parameters $\gamma$
and $\beta$ from the weak-field regime, we must have $q_V\simeq 1$.
In particular, for the minimal deviation from Schwarzschild given by $q_V=1$, we have
\be
\rh
=
\frac{6\,\gn\,M}{3\,\sqrt{3}-1}
\simeq
1.43\,\gn\,M
\ ,
\ee
which is also significantly larger than the corresponding harmonic Schwarzschild radius $r_{\rm S}$.
\par
Since $\Rh/\rh\sim 1$, the perturbative PPN expansion~\eqref{expWg}
cannot be effectively extended into the near-horizon region.
We instead have
\be
B
=
1+2\,V_0
=
\left(1-\frac{\rh}{r}\right)
\mathcal B
\ ,
\label{sB}
\ee
where $\mathcal B=\mathcal B(r)$ is a regular and strictly positive function for $r\ge \rh$,
which can be Taylor expanded as
\be
\mathcal B
=
\sum_{k=0}
\beta_k\left(\frac{r-\rh}{\rh}\right)^k
\ .
\label{BB}
\ee
Of course, the coefficients $\beta_k$ are fully determined from the explicit form of $V_0$, although their
expressions are rather cumbersome.
The first few ones, for instance, are given by
\be
\begin{split}
&
\beta_0
=
\frac{{(1+2\,q_V)^{3/2}-1}}{3\,q_V\,\sqrt{1+2\,q_V}}
\simeq
0.81
\\
&
\beta_1
=
\frac{q_V \left(3+6\,q_V+4\, q_V^2\right)
-(1+2\,q_V)^{3/2}}{9\, q_V\,(1+2\,q_V)^2}
\simeq
0.11
\\
&
\beta_2
\simeq
-0.07
\ ,
\end{split}
\ee
where the numerical estimates are obtained by setting $q_V=1$. 
\par
In order to change to the standard coordinates, we similarly expand the harmonic coordinate $r$
around $\bar r_{\rm H}\equiv \bar r(\rh)$ as
\be
r
=
\rho_0\,\bar r_{\rm H}
+
\bar r_{\rm H}\,
\sum_{k=1} \rho_k \pqty{\frac{\bar r-\bar r_{\rm H}}{\bar r_{\rm H}}}^k
\ ,
\label{rBbr}
\ee
where $\rh=\rho_0\,\bar r_{\rm H}$ is again the harmonic horizon radius in Eq.~\eqref{rhN}.
By inserting Eq.~\eqref{rBbr} into Eq.~\eqref{sB}, one can write 
\be
\bar B
=
\left(1-\frac{\bar r_{\rm H}}{\bar r}\right)
\bar {\mathcal B}
\ ,
\ee
with
\be
\bar{\mathcal B}
=
\sum_{k=0} 
{\mathcal B}_k 
\pqty{\frac{\bar r-\bar r_{\rm H}}{\bar r_{\rm H}}}^k
\ ,
\ee
where the coefficients ${\mathcal B}_k$ are determined by the known $\beta_j$'s in Eq.~\eqref{BB}
and the still undetermined $\rho_j$'s in Eq.~\eqref{rBbr}.
We notice in particular that $\bar B(\bar r>\bar r_{\rm H})>0$ implies that ${\mathcal B}_0>0$
and each ${\mathcal B}_k$ depends on the $\rho_{j\le k+1}$'s, which quickly makes
all expressions very cumbersome.
\par
In order to have a proper event horizon, we must require that
both $\bar B$ and $\bar A$ become negative for $\bar r<\bar r_{\rm H}$.
We thus assume~\footnote{Note that we require that the determinant
of the metric $g\sim \bar A\,\bar B$ is regular everywhere for $\bar r\ge \bar r_{\rm H}$.}
\be
\bar A
=
\left(1-\frac{\bar r_{\rm H}}{\bar r}\right)^{-1}
\bar{ \mathcal  A}
\ ,
\ee
where the function $\bar{ \mathcal  A}$ 
is also regular and strictly positive for $\bar r\ge\bar r_{\rm H}$ and can be
expanded as
\be
\bar{ \mathcal  A}
=
\sum_{k=0} {\mathcal A}_k \pqty{\frac{\bar r-\bar r_{\rm H}}{\bar r_{\rm H}}}^k
\ ,
\ee
where ${\mathcal A}_0>0$.
It follows that $\bar C=\bar{\mathcal A}\,\bar{\mathcal B}$ and, upon replacing into
Eq.~\eqref{equation for harmonic coordinate second form}, we obtain
\be
\bar r\,r''
-
\frac{2\,\bar{\mathcal A}\,r}{\bar r-\bar{r}_{\rm H}}
+
\left(
2
+
\frac{\bar{r}_{\rm H}}{\bar r-\bar{r}_{\rm H}}
+
\frac{\bar r\,\bar{\mathcal B}'}{2\,\bar{\mathcal B}}
-
\frac{\bar r\,\bar{\mathcal A}'}{2\,\bar{\mathcal A}}
\right)
r'
=
0
\ ,
\label{harmonicH}
\ee
where primes again denote derivatives with respect to $\bar r$.
In principle, this equation can be solved order by order in $(\bar r-\bar{r}_{\rm H})$,
thus relating the coefficients ${\mathcal A}_k$ to the ${\mathcal B}_k$'s and
$\rho_k$'s (equivalently, to the $\beta_k$'s and $\rho_k$'s).
\par
At leading order, for $\bar r\simeq \bar{r}_{\rm H}$, we have
\be
\frac{\bar{r}_{\rm H}}
{\bar r-\bar{r}_{\rm H}}
\left(
\rho_1
-
2\,\rho_0\,{\mathcal A}_0
\right)
\simeq
0
\ ,
\label{HA0}
\ee
which implies
\be
{\mathcal A}_0
=
\frac{\rho_1}{2\,\rho_0}
\ .
\label{A0}
\ee
At next to leading order, we then have 
\be
\rho_2
&\!\!=\!\!&
\rho_0\,{\mathcal A}_1
+\rho_1\,{\mathcal A}_0
-
\frac{\rho_1}{4}
\left(
4
+
\frac{{\mathcal B}_1}{{\mathcal B}_0}
-
\frac{{\mathcal A}_1}{{\mathcal A}_0}
\right)
\nonumber
\\
&\!\!=\!\!&
\frac{\rho_0}{2}
\left[
3\,{\mathcal A}_1
-
{\mathcal A}_0
\left(
4-4\,{\mathcal A}_0
+
\frac{{\mathcal B}_1}{{\mathcal B}_0}
\right)
\right]
\ ,
\label{HA00}
\ee
where we recall that ${\mathcal A}_0$ and ${\mathcal B}_0$ must be positive.
In particular, if $\rho_1\simeq 1$ and $|\rho_2|\ll 1$,~\footnote{We will see next that this
is a rather accurate estimate.} we must have
\be
{\mathcal A}_0
\simeq
\frac{1}{2\,\rho_0}
\ ,
\qquad
{\mathcal A}_1
\simeq
\frac{2}{3\,\rho_0}
\left(
1
-\frac{1}{2\,\rho_0}
-\frac{\mathcal B_1}{4\,\mathcal B_0}
\right)
\ ,
\ee
where the known and exact coefficient 
\be
{\mathcal B}_0
=
\frac{\beta_0}{\rho_0}
=
\frac{(1+2\,q_V)^{3/2}-1}
{3\,\rho_0\,q_V\,\sqrt{1+2\,q_V}}
\ ,
\label{B0}
\ee
and, since ${\mathcal B}_1$ depends also on $\rho_2$, we do not show its rather long expression
here.
\par
It is important to remark that the unknown coefficients
${\mathcal A}_k$'s depend on the coefficients $\rho_k$'s, both through
Eq.~\eqref{harmonicH} and because the ${\mathcal B}_k$'s depend
on the $\rho_k$'s. 
The only way to fix this ambiguity, related with the expression of the
harmonic $r=r(\bar r)$, on physical grounds is to match the near-horizon expressions 
of the metric components $\bar A$ and $\bar B$ with their analogue in the weak-field
regime.
The latter was obtained previously by imposing observational
constraints to partly fix $r=r(\bar r)$ therein.
The matching between the two regimes will therefore leave unspecified only those
parameters which do not conflict with the experimental bounds at large distance
from the source.
\subsection{Matching with weak field}
Let us start from noting that the Taylor expansion for the near-horizon regime
is comparable with the one for weak field when
\be
\frac{\bar r - \bar r_{\rm H}}{\bar r_{\rm H}}
\simeq
\frac{\Rh}{\bar r}
\ ,
\ee
or $\bar r\simeq \bar r_{\rm m}$, with
\be
\bar r_{\rm m}
&\!\!=\!\!&
\frac{\bar r_{\rm H}}{2}
\left(
1+\sqrt{1+4\,\frac{\Rh}{\bar r_{\rm H}}}
\right)
\nonumber
\\
&\!\!=\!\!&
\frac{\rh}{2\,\rho_0}
\left(
1+\sqrt{1+4\,\rho_0\,\frac{\Rh}{\rh}}
\right)
\ ,
\label{defrm}
\ee
where we recall that $\rho_0>0$ and the harmonic $\rh$ is given in Eq.~\eqref{rhN}.
Moreover, the first few terms in the two expansions still provide a reliable approximation
at $\bar r=\bar r_{\rm m}$ if 
\be
\frac{\Rh}{\bar r_{\rm m}}
=
2\,\rho_0\,\frac{\Rh}{\rh}
\left(
1+\sqrt{1+4\,\rho_0\,\frac{\Rh}{\rh}}
\right)^{-1}
\lesssim
1
\ .
\ee
The above condition is satisfied for
\be
\rho_0
\lesssim
\rho_{\rm c}
\equiv
2\,\frac{\rh}{\Rh}
=
\frac{6\,q_V}{(1+2\,q_V)^{3/2}-1}
\ .
\ee
\par
In particular, by matching the expressions of the harmonic coordinates~\eqref{expWr} and~\eqref{rBbr}
at $\bar r=\bar r_{\rm m}$, that is
\be
\label{matching}
\bar r_{\rm m}
-
\frac{\Rh}{2}
+
\bar r_{\rm m}\,
\sum_{k=2} \sigma_{k} 
\left(
\frac{\Rh}{\bar r_{\rm m}}
\right)^k
&\!\!=\!\!&
\rho_0\,\bar{r}_{\rm H}
+
\bar{r}_{\rm H}\,
\sum_{k=1} 
{\rho_k} 
\left(
\frac{\bar r_{\rm m}-\bar r_{\rm H}}{\bar r_{\rm H}}
\right)^k
\nonumber
\\
&\!\!=\!\!&
\rho_0\,\bar{r}_{\rm H}
+
\rho_1\left(
\bar r_{\rm m}
-\bar r_{\rm H}
\right)
+
\bar r_{\rm H}\,
\sum_{k=2} 
{\rho_k} 
\left(
\frac{\Rh}{\bar r_{\rm m}}
\right)^k
\ ,
\ee
we obtain
\be
\rho_0
=
(1-\rho_1)\,\frac{\bar r_{\rm m}}{\bar r_{\rm H}}
+
\rho_1
-
\frac{\Rh}{2\,\bar r_{\rm H}}
-
\sum_{k=2}
\left(\rho_k-\frac{\bar r_{\rm m}}{\bar r_{\rm H}}\,\sigma_k\right)
\left(
\frac{\Rh}{\bar r_{\rm m}}
\right)^k
\ .
\label{rho0m}
\ee
At leading order, we thus find
\be
\rho_0
\simeq
(1-\rho_1)\,\frac{\bar r_{\rm m}}{\bar r_{\rm H}}
+
\rho_1
-
\frac{\Rh}{2\,\bar r_{\rm H}}
\ .
\label{rho0_1}
\ee
This estimate can be further improved by considering yet another expansion
about $\bar r =\bar r_{\rm m}$ and determining the corresponding Taylor coefficients
from the matching with the weak-field expansion for $\bar r\gtrsim \bar r_{\rm m}$
and with the near-horizon expansion for $\bar r\lesssim \bar r_{\rm m}$.
This is equivalent to imposing continuity of the function $r=r(\bar r)$ and
its derivatives across $\bar r_{\rm m}$ (see Appendix~\ref{A:direct}).
We remark here that the result for $|c_2|=|\xi-1|\lesssim 1$ is consistent with the above
expressions for $\rho_1\simeq 1$ and $|\rho_2|\ll 1$.
\subsection{Near-horizon geometry}
The better estimate of $\rho_0$ in Eq.~\eqref{rho_} yields for the areal radius of the
bootstrapped Newtonian horizon
\be
\bar r_{\rm H}
=
\frac{\rh}{\rho_0}
\simeq \pqty{1.21 +0.27\, c_2}
\Rh
\ .
\ee
The value of $c_2=c_2^{\rm S}$ which would give $\bar r_{\rm H}=\Rh=2\,\gn\,M$ according
to this equation is outside our range of approximation (namely, $|c_2|\ll 1$).
In fact, resorting to Eq.~\eqref{rho0full}, we obtain $c_2^{\rm S}\approx -0.696$, corresponding
to $\xi=c_2^{\rm S}+1\simeq 0.3$.
\par
\begin{figure}[t!]
\centering
\includegraphics[width=8cm]{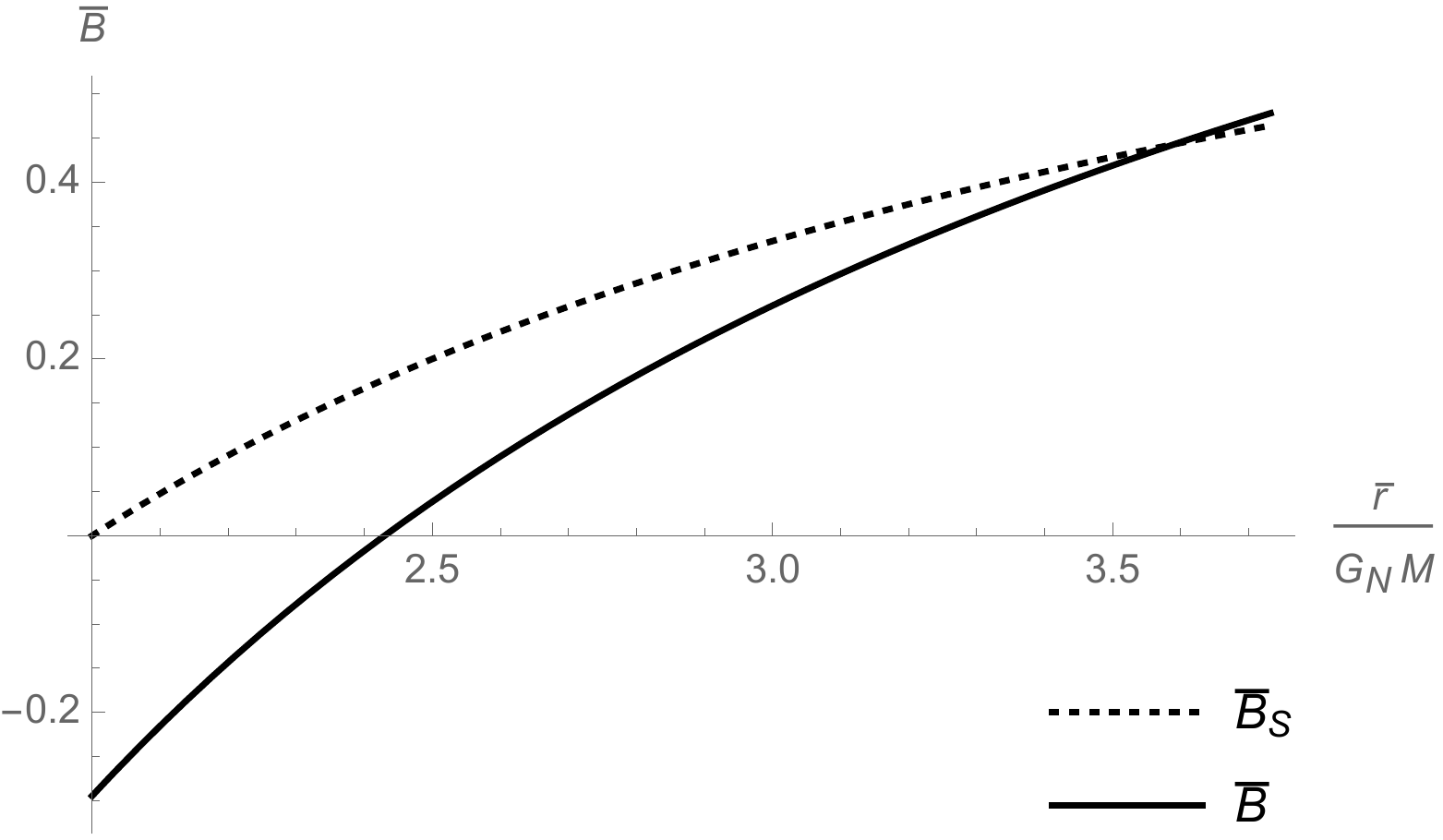}
$\ $
\includegraphics[width=8cm]{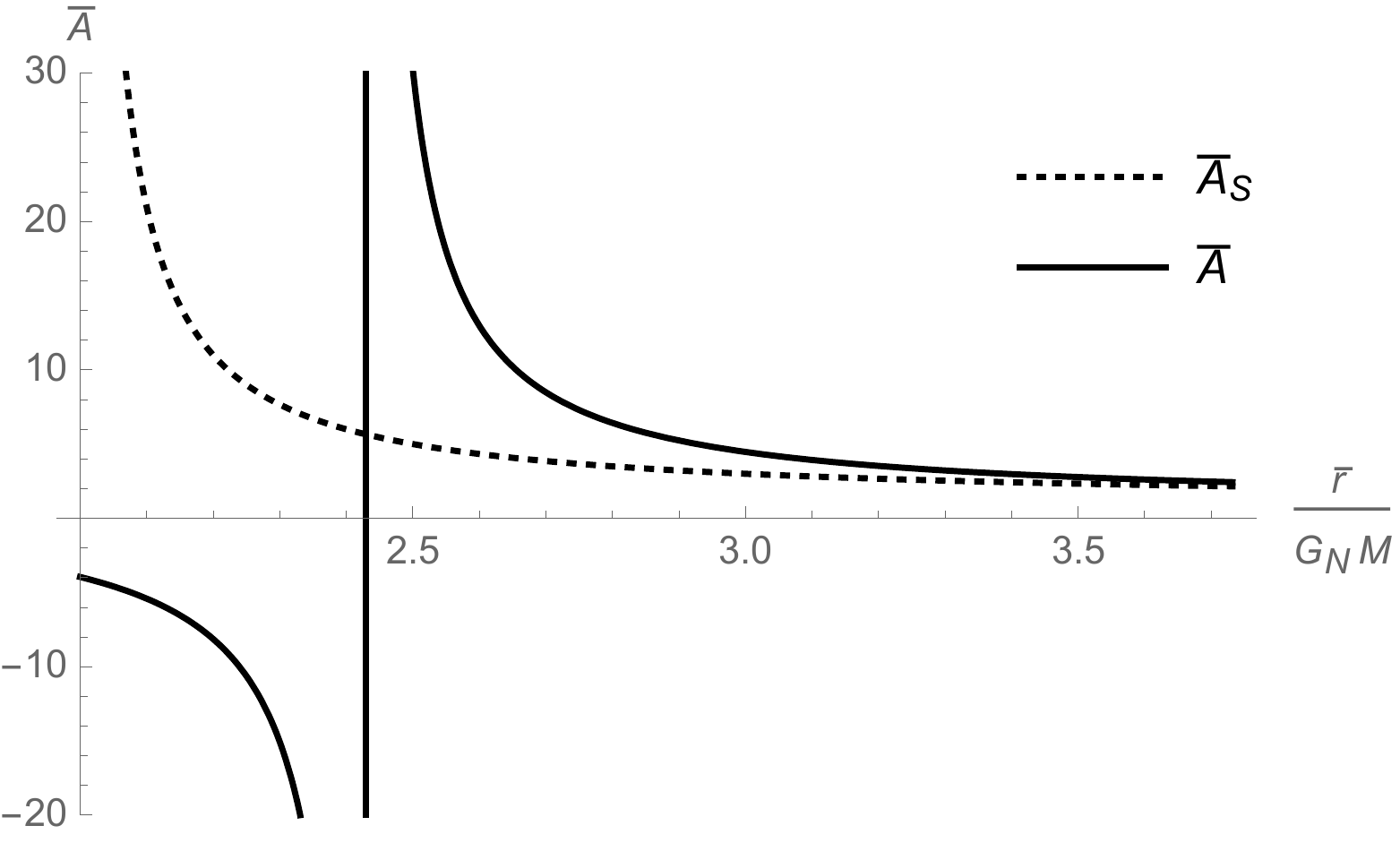}
\caption{Comparison between the Schwarzschild and bootstrapped Newtonian 
metric components for $\Rh<\bar r<\bar{r}_{\rm m}$.
The vertical line in the right panel is the location of the bootstrapped horizon $\bar r=\bar{r}_{\rm H}$.}
\label{f:match}
\end{figure}
\begin{figure}[b]
\centering
\includegraphics[width=8cm]{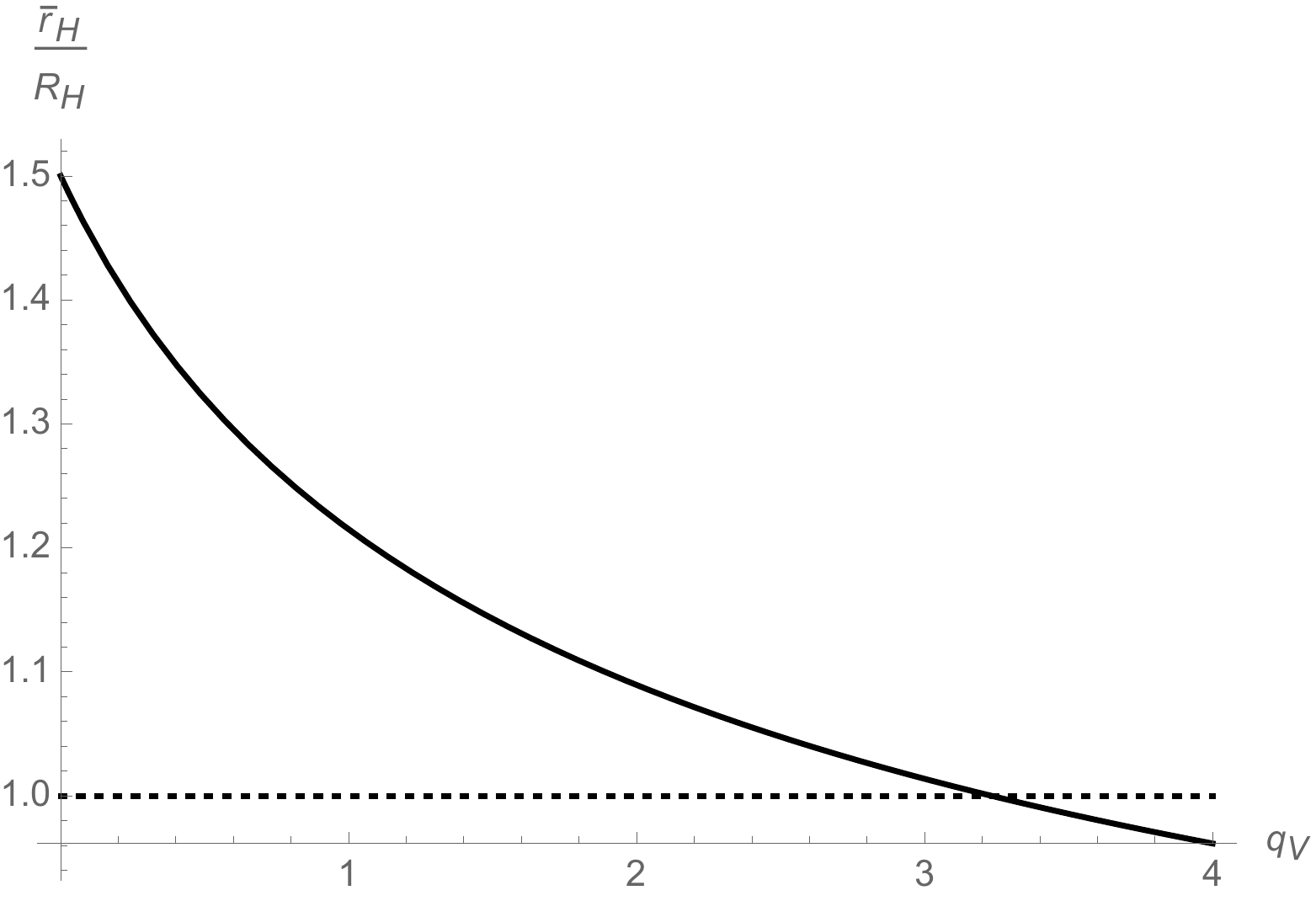}
\caption{Bootstrapped horizon $\bar{r}_{\rm H}$ in units of $\Rh$.
The horizontal dotted line is unity.}
\label{f:rHh}
\end{figure}

On using Eqs.~\eqref{rho_}, \eqref{B0} and \eqref{A0}, we obtain
\be
\begin{split}
{\mathcal B}_0
&
\simeq
1.37 + 0.50\, c_2
\\
{\mathcal A}_0
&
\simeq
0.85 + 0.31\, c_2
\ .
\end{split}
\ee
In particular, for $c_2=0$, we find $\rho_1\simeq 1$ and Eq.~\eqref{rho0_1} yields
the same relation between the harmonic and the areal horizon radii which holds
for the Schwarzschild solution, that is
\be
\bar r_{\rm H}
\simeq
\rh
+
\frac{\Rh}{2}
\ .
\ee
From the bootstrapped potential we thus obtain
\be
\bar r_{\rm H}
\simeq
\frac{(1+2\,q_V)^{3/2}-1+6\,q_V}{(1+2\,q_V)^{3/2}-1}
\,\gn\,M
\simeq
2.43\,\gn\,M
\ ,
\label{brh}
\ee
where the last value is for $q_V=1$.
The corresponding metric coefficients $\bar B$ and $\bar A$ at leading order in the near-horizon
expansion are shown in Fig.~\ref{f:match}, where they are compared with their Schwarzschild analogues.
The only relevant difference is given by the areal radius of the bootstrapped horizon.
For this reason we plot $\bar{r}_{\rm H}$ in units of $\Rh$ in Fig.~\ref{f:rHh}, and note that
$\bar{r}_{\rm H}=\Rh$ for
\be
q_V
=
\frac{3+2\,\sqrt{3}}{2}
\simeq
3.23
\ .
\ee
Clearly, this much stronger self-coupling would not be compatible with the weak-field bounds,
further supporting the result that the bootstrapped Newtonian metric contains an horizon
$\bar{r}_{\rm H}$ larger than Schwarzschild's $\Rh$.
\par
Since the matching radius $\bar r_{\rm m}\simeq 3.73\,\gn\,M$, one can expect a
correction for the radius $\bar r_{\rm ph}$ of the photon orbit, whose value is $3\,\gn\,M$
in General Relativity.
Using $\bar C\simeq {\mathcal A}_0\,{\mathcal B}_0\simeq 1.17$ and constant,
the latter can be estimated from Eq.~\eqref{GeoRa} as
\be
0
=
\mathcal V_{\rm eff}'
\sim
3\,\bar r_{\rm H}
-2\,\bar r_{\rm ph}
\ ,
\ee
where $\mathcal V^{\rm eff}$ is the potential in Eq.~\eqref{VeffR} with $k=0$
for null trajectories.
The result $\bar r_{\rm ph}\simeq 3.64\,\gn\,M$ is just short of $\bar r_{\rm m}$,
and a better reconstruction of the near-horizon metric including a few higher order
terms ${\mathcal A}_k$ and ${\mathcal B}_k$ is therefore likely to modify this value.
In fact, we note that $\bar C$ must approach the General Relativistic value
$\bar C=1$ rather fast in the weak-field regime, according to Eq.~\eqref{bCw}, and
$\bar C'$ cannot therefore be neglected near the horizon.
For example, if we simply employ a linear approximation for $\bar{\mathcal B}$,
and take $\rho_1=1$ and $\rho_2=0$, we get
\be
\bar r_{\rm ph}
\simeq
\frac{3\,\mathcal B_0-2\,\mathcal B_1}
{2\,\mathcal B_0-2\,\mathcal B_1}\,
\bar r_{\rm H}
\simeq 
3.26\,\gn\,M
\ ,
\ee
which is closer to the prediction of General Relativity.
\par
On the other hand, the innermost stable circular orbit of General Relativity is located
at $\bar r_{\rm ISCO}=6\,\gn\,M$, and its location in the bootstrapped Newtonian metric
should instead be recovered rather accurately from the weak-field approximation.
From Eq.~\eqref{Bc2} and~\eqref{Ac2} evaluated at $\bar r=\bar r_{\rm ISCO}$ we indeed
obtain for the deviation of the bootstrapped metric from Schwarzschild's
\be
\begin{split}
&
\frac{\bar B-\bar B_{\rm S}}
{\bar B_{\rm S}}
\simeq
\frac{5+c_2}{216}
\simeq
0.02
\\
&
\frac{\bar A-\bar A_{\rm S}}
{\bar A_{\rm S}}
\simeq
\frac{5+17\,c_2}{72}
\simeq
0.07
\ ,
\end{split}
\ee
where we expect $|c_2|=|\xi-1|\ll 1$.
\subsection{Harmonic and areal compactness}
For a source of harmonic radius $R$ in the Schwarzschild space-time, one can introduce the
compactness in the harmonic coordinate as
\be
X_{\rm S}
\equiv
\frac{\gn\,M}{R}
\ ,
\ee
or in the areal coordinate as
\be
\bar X_{\rm S}
\equiv
\frac{2\,\gn\,M}{\bar R}
=
\frac{2\,X_{\rm S}}{1+X_{\rm S}}
\ ,
\ee
where we used Eq.~\eqref{rSbr}.
In particular, $\bar X_{\rm S}(\Rh)=X_{\rm S}(r_{\rm S})=1$ for a Schwarzschild black hole.
\par
For the bootstrapped metric, we could likewise introduce the harmonic compactness
\be
X
\equiv
\frac{\rh}{R}
\ee
and the areal compactness
\be
\bar X
\equiv
\frac{\bar r_{\rm H}}{\bar R}
=
\frac{X}{\rho_0+(1-\rho_0)\,X}
\ ,
\ee
in which we employed the leading order transformation~\eqref{rBbr} with $\rho_1\simeq 1$,
\be
r
\simeq
\rho_0\,\bar r_{\rm H}
+\bar r -\bar r_{\rm H}
\ ,
\ee
so that $\bar X(\bar r_{\rm H})=X(\rh)=1$.
\par
For the purpose of comparing with General Relativity, it is however more convenient
to use the Schwarzschild quantities and note that, for a bootstrapped Newtonian black hole
\be
X_{\rm S}(\rh)
=
\frac{\gn\,M}{\rh}
=
\frac{6\,q_V}{(1+2\,q_V)^{3/2}-1}
\simeq
0.70
\ee
and
\be
\bar X_{\rm S}(\bar r_{\rm H})
=
2\,\rho_0\,X_{\rm S}
=
\frac{2\,X_{\rm S}}{1+X_{\rm S}}
\simeq
0.83
\ ,
\ee
where the numerical values are those for $q_V=1$, as usual.
We notice incidentally that this value is just slightly smaller than the Buchdahl limit
$\bar X_{\rm B}=8/9\simeq 0.89$. 
\section{Conclusions and outlook}
\label{S:conc}
\setcounter{equation}{0}
The bootstrapped Newtonian approach is devised to capture quantum effects
which induce large (mean-field) deviations from classical General Relativity when large 
matter sources are involved.
Such effects would be completely determined if we knew the proper quantum state
describing specific self-gravitating systems.
What we know for certain is that the strong field regime of gravity governed by the
Einstein field equations is not linear.
Determining the relevant quantum state therefore requires that one solves nonlinear quantum
dynamics, which seems hardly a tenable task for large and very compact sources.
The bootstrapped Newtonian approach considers a simplified form of nonlinear 
dynamics for gravity, compared to General Relativity, but aims at including quantum
deviations from classicality in a form that is sufficiently general to confront observations.
This generality is manifested in the coupling constants appearing in the action~\eqref{LagrV}.
\par
The potential experienced by test particle at rest is however not sufficient
to determine all deviations from the classical solutions of General Relativity.
Starting from the bootstrapped Newtonian potential outside a static and
spherically symmetric source, we here obtained a complete metric by supplying
further conditions of compatibility with observations in the weak-field regime.
The main difference with respect to the unique Schwarzschild solution of
General Relativity.
is given by the larger horizon radius estimated in Eq.~\eqref{brh}.
This prediction makes the bootstrapped Newtonian programme experimentally
testable, for instance, by measurements of light trajectories reaching the photon orbit.
A more detailed analysis of these trajectories in terms of the parameters 
of the effective metric is the natural continuation of the work presented here.
\par
A possible conclusion of such a phenomenological analysis could
be that a consistent description of the near-horizon region of black holes requires
more than the first few nonlinear terms included in the bootstrapped Newtonian
Lagrangian~\eqref{LagrV}.
This possibility will be investigated in the future, but, in this respect, it is important
to recall that the entire programme about bootstrap Newtonian gravity is motivated
by the idea that black holes and similarly compact sources might require a fully quantum,
rather than semiclassical, description.
It is therefore not {\em a priori\/} clear to what extent the effective metric
we obtained is meaningful at such short distances from the (would-be classical)
horizon. 
More precisely, one expects that the interaction of matter and light falling towards
the black hole should be described in terms of scattering processes, for which
classical geodesic lines will become an unreliable approximation if black holes
are indeed extended quantum objects (for a non-exhaustive list, see
Refs.~\cite{DvaliGomez,ciarfella,nicolini,Dai:2020irc,Bojowald:2020dkb,
Casadio:2013hja,Calmet:2019eof,bonanno}).
This viewpoint will also require a more detailed quantum description of the
matter source itself, which is left completely out here.
\par
Finally, we would like to mention that the weak-field regime is also worthy of
further study.
First of all, there is the possibility that deviations from the Schwarzschild
geometry reproduce the kind of effective dark fluid responsible for Dark Matter
phenomenology as explored in Refs.~\cite{cosmo}.
Moreover, propagation of gravitational waves and other signals would also be affected
by the non-trivial background corresponding to the effective metric.
All of these developments are left for future investigations.
\subsection*{Acknowledgments}
R.C.~and I.K.~are partially supported by the INFN grant FLAG.
The work of R.C.~and A.G~has also been carried out in the framework
of activities of the National Group of Mathematical Physics (GNFM, INdAM)
and COST action {\em Cantata\/}. 
%
%
%
%%%%%%%%%%%%%%%%%%%%%%%%%%%%%%%%%%%%%%%%%%
%
\appendix
\section{Weak-field effective energy-momentum tensor}
\label{A:emt}
\setcounter{equation}{0}
The effective fluid density for general values of the Robertson-Eddington parameters is given by
\be
\rho^{\rm eff}
&\!\!\simeq\!\!&
\frac{\gn\,M^2}{4\, \pi\,  \bar r^4}
\left\{
\left(\beta -3\,\gamma+2\,\gamma^2 -2\,c_2\right)
\phantom{\frac{A}{B}}
\right.
\nonumber
\\
&&
\qquad
\left.
-
\left[
5+
32\, \beta ^2
-12\,\beta\,  (3- \gamma)
+\gamma\,(18-3\,\gamma-8\,\gamma^2)
+8\, c_2\, (2+\gamma)
\right]
\frac{\gn\,M }{2\, \bar r}
\right\}
\ ,
\ee
the pressure by
\be
p_r^{\rm eff}
&\!\!\simeq\!\!&
\frac{M}{4\, \pi\, \bar r^3}
\left\{(1-\gamma)
+
\left(
2-\beta-3 \,\gamma +2 \,\gamma^2 -2\, c_2
\right)
\frac{\gn\, M }{\bar r}
\right.
\nonumber
\\
&&
\qquad\quad
\left.
+
\left(1+\gamma\right)
\left(
1-3\,\gamma+2\,\gamma^2-2\,c_2
\right)
\frac{\gn^2\, M^2 }{\bar r^2}
\right\}
\ ,
\ee
and the tension by
\be
p_t^{\rm eff}
&\!\!\simeq\!\!&
\frac{M}{8\, \pi\, \bar r^3}
\left\{(\gamma-1)
+
\left(
2\,\beta-3
+5\,\gamma-4\,\gamma^2
+4\,c_2
\right)
\frac{\gn\, M }{\bar r}
\right.
\nonumber
\\
&&
\qquad\quad
\left.
+
\left[
(1+\gamma)
\left(
1-2\,\beta+\gamma+6\,\gamma^2
\right)
+c_2\,(2+6\, \gamma)
\right]
\frac{\gn^2\, M^2 }{\bar r^2}
\right\}
\ .
\ee
The anisotropy $\Pi\equiv p_r-p_t$ therefore is
\be
\Pi
&\!\!\simeq\!\!&
\frac{M}{8\, \pi\, \bar r^3}
\left\{
3\,(1-\gamma)
+
\left(
7 - 4\,\beta -11\,\gamma + 8\, \gamma^2  - 8\,c_2
\right)
\frac{\gn\, M }{\bar r}
\right.
\nonumber
\\
&&
\qquad\quad
\left.
+\left[
(1-\gamma)
\left(
1+2\,\beta-3\,\gamma+10\,\gamma^2
\right)
-2\,c_2\,(3+5\, \gamma)
\right]
\frac{\gn^2\, M^2 }{\bar r^2}
\right\}
\ .
\ee
The Misner-Sharp-Hernandez mass reads
\be
m(\bar r)
&\!\!\simeq\!\!&
M
\left\{
\gamma
-
\left(\beta-3\,\gamma+2\, \gamma^2-2\, c_2\right)
\frac{\gn\, M }{\bar r}
\right.
\nonumber
\\
&&
\qquad
\left.
+
\left[
5
-32\,\beta^2
-12\,\beta\,(3-\gamma)
+18\,\gamma-3\,\gamma^2-8\,\gamma^3
+8\,c_2 \,(2+\gamma)
\right]
\frac{\gn^2\, M^2 }{4\,\bar r^2}
\right\}
\ .
\ee
For $\beta=\gamma=1$, the above expressions reduce to those shown in the main
text.
\section{Intermediate range expansion}
\label{A:direct}
\setcounter{equation}{0}
Let us start by expanding the Schwarzschild metric in harmonic coordinates
around the horizon $r_{\rm S}=\Rh/2$.
From the general form~\eqref{harmonic form} and Eq.~\eqref{rSbr} we obtain
\be
B_{\rm S}
=
\frac{1}{A_{\rm S}}
\simeq
\frac{r-\gn\,M}{2\,\gn\,M}
\ .
\ee
The analogous expansion for the coefficient $B$ of the bootstrapped Newtonian metric
can be derived from Eq.~\eqref{sB}.
To be more specific, we shall only consider the case $q_V=\gamma=1$ here, which yields
\be
B
\simeq \frac{r-1.43\,\gn\,M}{1.77\,\gn\,M}
\ .
\ee
However, we need both $\rho_0$ and $\rho_1$ in Eq.~\eqref{rBbr} in order to obtain the leading order 
expression for the coefficient $A$.
\par
For this purpose, we expand $r=r(\bar r)$ around $\bar r_{\rm m}$ defined in Eq.~\eqref{defrm}
as the radius at which the weak-field expansion becomes comparable to the near-horizon one.
This intermediate expansion of $r=r(\bar r)$ can then be constrained by using the weak-field expansion
to the right of $\bar r_{\rm m}$ and the near-horizon expansion to the left of $\bar r_{\rm m}$.
In particular, continuity of $r=r(\bar r)$ and of its first few derivatives around $\bar r_{\rm m}$ implies
\be
\bar r_{\rm H}
\bqty{\rho_0+\rho_1\,\frac{\Rh}{\bar r_{\rm m}}+\rho_2\pqty{\frac{\Rh}{\bar r_{\rm m}}}^2+\mathcal O(3)}
&\!\!=\!\!&
\bar r_{\rm m}
\bqty{1-\frac{\Rh}{2\,\bar r_{\rm m}}-\frac{c_2}{2}\pqty{\frac{\Rh}{\bar r_{\rm m}}}^2+\mathcal O(3)}
\\
\rho_1+2\,\rho_2\,\frac{\Rh}{\bar r_{\rm m}}
+\mathcal O(2)
&\!\!=\!\!&
1+\frac{c_2}{2}\pqty{\frac{\Rh}{\bar r_{\rm m}}}^2
+\mathcal O(3)
\\
\frac{1}{\bar r_{\rm H}}
\bqty{2\,\rho_2+\mathcal O(1)}
&\!\!=\!\!&
\frac{1}{\bar r_{\rm m}}
\bqty{-c_2\pqty{\frac{\Rh}{\bar r_{\rm m}}}^2+\mathcal O(3)}
\ ,
\ee
where $\mathcal O(k)$ denotes a quantity proportional to $\pqty{\Rh/\bar r_{\rm m}}^k$.
Solving these equations gives
\be
\label{rho0full}
\rho_0
&\!\!=\!\!&
1-\frac{\Rh}{2\,\bar r_{\rm m}}
-\frac{c_2+1}{2}\pqty{\frac{\Rh}{\bar r_{\rm m}}}^2
-\frac{c_2}{2}\pqty{\frac{\Rh}{\bar r_{\rm m}}}^3
+\mathcal O(3)
\\
\rho_1
&\!\!=\!\!&
1+\frac{c_2}{2}\pqty{\frac{\Rh}{\bar r_{\rm m}}}^2
+\mathcal O(2)
\\
\rho_2
&\!\!=\!\!&
\mathcal O(1)
\ .
\ee
We next note that 
\be
\rho_0
=
\frac{\rh}{\bar r_{\rm H}}
=
\frac{\rh}{\Rh}\,
\frac{\Rh}{\bar r_{\rm m}}\,
\frac{\bar r_{\rm m}}{\bar r_{\rm H}}
=
\frac{\rh}{\Rh}\,
\frac{\Rh}{\bar r_{\rm m}}
\left(
\frac{\Rh}{\bar r_{\rm m}}
+1
\right)
\ ,
\ee
where the bootstrapped Newtonian horizon $\rh$ is given in Eq.~\eqref{rhN} and the last equality follows
from the definition~\eqref{defrm}. 
With this expression for $\rho_0$, Eq.~\eqref{rho0full} can be used to relate ${\Rh}/{\bar r_{\rm m}}$
to the weak-field coefficient $c_2=\xi-1$, and one can then obtain explicit estimates for $\rho_0$, $\rho_1$
and $\rho_2$.
Using again $q_V=1$, we get
\be
\label{Rhtorm}
\frac{\Rh}{\bar r_{\rm m}} \simeq 0.54 - 0.09\, c_2+\mathcal O(3)
\ ,
\ee
which is smaller than one, as required for the validity of the truncation.
Correspondingly, we have
\be
\begin{split}
\label{rho_}
&
\rho_0
\simeq
0.59-0.13\,c_2+\mathcal O(3)
\\
&
\rho_1
\simeq 
1+0.14\,c_2+\mathcal O(2)
\ .
\end{split}
\ee
From Eq.~\eqref{Rhtorm}, we can further estimate the orders of magnitude of neglected 
quantities, namely $\mathcal O(1)\approx 0.54$, $\mathcal O(2)\approx 0.29$ and
$\mathcal O(3)\approx 0.15$, assuming proportionality constants of order one as well.
\par
Moreover, using the above estimates in Eq.~\eqref{rho0_1} yields $\rho_0\simeq 0.59$ to leading
order.
This confirms that the direct matching between the weak-field and near-horizon expansions
is already rather accurate.
In fact, the ratio between the first term that we neglected and the last we included in the direct
matching in Eq.~\eqref{matching}, that is
\be
\abs{\frac{\rho_2\,\bar r_{\rm H}-\sigma_2\,\bar r_{\rm m}}
{\rho_1\,\bar r_{\rm H}-\sigma_1\,\bar r_{\rm m}}\cdot \frac{\Rh}{\bar r_{\rm m}}}
\simeq
0.08+0.19\, c_2
\ ,
\ee
is reasonably small in the expected range of values of $c_2=\xi-1$.
\par
Finally, Eq.~\eqref{A0} with the above estimates yields
\be
A
\simeq \frac{\pqty{2.06+1.50 \, c_2}\,\gn\,M}{r-1.43\,\gn\,M}
\ ,
\ee
where we used the approximation of small $c_2$ and neglected all $\mathcal O(k)$ terms.
\end{document}